\documentclass{aastex62}
\usepackage{hyperref}

\graphicspath{{./}{figures/}}

\accepted{June 18, 2019 - The Astrophysical Journal}

\shorttitle{The intermediate polar systems TV\,Col and V2731\,Oph}
\shortauthors{Lopes de Oliveira \& Mukai}

\begin{document}

\title{DEVELOPING THE PHYSICAL UNDERSTANDING OF INTERMEDIATE POLARS:\\ AN X-RAY STUDY OF TV COL AND V2731 OPH}

\correspondingauthor{Raimundo Lopes de Oliveira}
\email{rlopes@ufs.br}

\author{R. Lopes de Oliveira}
\affiliation{X-ray Astrophysics Laboratory, NASA Goddard Space Flight Center, Greenbelt, MD 20771, USA}
\affiliation{Center for Space Science and Technology, University of Maryland, Baltimore County, 1000 Hilltop Circle, Baltimore, MD 21250, USA}
\affiliation{Departamento de F\'isica, Universidade Federal de Sergipe, Av. Marechal Rondon, S/N, 49000-000 S\~ao Crist\'ov\~ao, SE, Brazil
}
\affiliation{Observat\'orio Nacional, Rua Gal. Jos\'e Cristino 77, 20921-400, Rio de Janeiro, RJ, Brazil}
\author{K. Mukai}
\affiliation{CRESST II and X-ray Astrophysics Laboratory, NASA Goddard Space Flight Center, Greenbelt, MD 20771, USA}
\affiliation{Center for Space Science and Technology, University of Maryland, Baltimore County, 1000 Hilltop Circle, Baltimore, MD 21250, USA}

\begin{abstract}

X-rays in intermediate polars (IPs) originate in a compact region near the surface of a magnetic white dwarf (WD) and interact with the complex environment surrounding the emission region. Here we report a case study of two IPs, TV\,Col and V2731\,Oph with selected archival X-ray observations (NuSTAR, Swift, Suzaku, and XMM-{\it Newton}). For TV\,Col, we were successful in simultaneously accounting for the primary X-rays, the secondary X-rays due to Compton scattering and fluorescence, and the effects of local absorbers. In this case, we were able to demonstrate that the shock height is small, based on the high reflection amplitude, and hence the maximum temperature of the post-shock region can be used to derive the WD mass of 0.735$\pm$0.015\,M$_\odot$. 
Despite the high specific accretion rate required to explain the small shock height, we do not
detect any spin modulation in our NuSTAR data, consistent with the modest amount of complex
absorption seen spectroscopically.
We argue that our results are robust because they are based on the joint temporal-spectral analysis of broadband X-ray data.  The spectrum of V2731\,Oph is more highly absorbed. Through our analysis of the Suzaku data, we present a spectral model with nitrogen overabundance without the previously claimed soft blackbody that should be further explored. We have been unable to constrain reflection amplitude for V2731\,Oph; this and the detection of spin modulation above 10\,keV suggest that it may have a tall shock, hence we only derive a lower limit to the mass of its WD ($>$ 0.9\,M$_{\odot}$).

\end{abstract}





\section{Introduction} \label{sec:intro}

Intermediate Polars (IPs) are a subgroup of Cataclysmic Variables (CVs) in which a moderately magnetic (10$^{5-7}$\,G) white dwarf (WD) accretes matter via Roche-lobe overflow from a companion, usually a low-mass star on or near the main sequence. An accretion disk is usually formed and its inner part is truncated by the magnetic field, which channels the gas in accretion columns towards the magnetic poles of the WD. Strong shocks are formed in the columns, heating up the gas to a high temperature defined in the first order by the depth of gravitational potential well of the WD ($k$T\,$\propto$\,M$_{WD}$/R$_{WD}$). The shocked gas subsequently settles onto the WD surface as it is cooled by radiating X-rays and cyclotron emission. Thus the X-ray emission is intrinsically multi-temperature in nature. 

Information on the accretion geometry and the white dwarf mass is encoded in the intrinsic emission and its subsequent interaction with its surroundings. The observed radiation in IPs has a significant contribution from X-ray photons that have interacted with the environment, including the stellar surface of the WD itself, the accretion column, the accretion disk, and any circumstellar medium that might exist. 
Part of the interacting X-rays suffers photoelectric absorption inducing a deficit in the observed X-rays which is more pronounced for softer photons. Another part is reflected and results in excess in the continuum around 10--30\,keV as a Compton hump and in fluorescent emission lines, notably the Fe\,K line at 6.4\,keV. 
The description of both components, absorption and reflection, is necessary for an in-depth understanding of the accretion geometry and characteristics of the intrinsic X-rays, which can also lead to the determination of the WD mass. This task requires wide-bandpass spectroscopy covering both soft and hard X-rays -- especially because the partial-covering absorption, expected in IPs due to their complex environment, can have a similar effect on the spectral shape around the Fe K edge as reflection.

The advent of imaging hard X-ray observations using NuSTAR has allowed the detailed investigation of X-ray spectra of IPs, including both absorption and reflection \citep{2015ApJ...807L..30M}. While other recent papers have concentrated on deriving the white dwarf mass using hard X-ray spectroscopy \citep{2018MNRAS.476..554S}, including the effects of tall shocks and small inner radius of the disk \citep{2019MNRAS.482.3622S}, concerns remain that such investigations can be compromised if inadequate attention is paid to the complex interplay of absorption, reflection, and the intrinsic spectral shape. Here we present our detailed analysis of wide band X-ray spectra of two IPs, TV\,Col and V2731\,Oph, to test the potential degeneracy in spectral analysis results. We also follow \citet{2015ApJ...807L..30M} in analyzing the hard X-ray spin modulation as a method to break the degeneracy.

\section{The targets}

According to GAIA DR2 parallax, TV\,Col is at a distance of 513$^{+4}_{-5}$\,pc 
\citep{2016A&A...595A...1G,2018arXiv180409365G}. The system has been investigated with many X-ray observatories since its first detection with the Ariel V satellite \citep[2A 0526-328;][]{1978MNRAS.182..489C}.
EXOSAT observations ($\sim$\,2--10\,keV) revealed it to be an absorbed hard X-ray emitter with flux modulated at 1,911\,s, interpreted as the white dwarf spin period \citep{1987Ap&SS.130..261S}. Based on the energy-dependent pulse profile from EXOSAT data, \citet{1989MNRAS.237..853N} argued for the presence of a complex absorption structure that could be explained by a partial-covering absorption. 
The spin period was refined to 1,909.7$\pm$2.5 s by \citet{2004AJ....127..489R} from a nearly sinusoidal modulation by using RXTE, ROSAT, and ASCA data, 
which also showed the 5.5\,h orbital modulation. 
The authors also claimed the presence of a strong attenuation of the X-rays and signals of partial-covering absorption from hardness ratio variations, and spin modulation decreasing with energy. Broadband RXTE spectra (PCA and HXTE; $\sim$\,3--100\,keV) confirmed that TV\,Col is a hard, thermal, X-ray source presenting a clear Fe\,K$\alpha$ complex \citep{2005A&A...435..191S}.
A single temperature thermal fit of the high energy BAT/Swift (15--195\,keV) data resulted in a temperature of $k$T\,=\,21.6$\pm$2.4\,keV \citep{2009A&A...496..121B}. An analysis of the Suzaku data in the 3--50\,keV range, with the inclusion of a partial-covering absorber, indicated a shock temperature $k$T\,=\,45.7$^{+16.6}_{-9.1}$\,keV in a plasma with a subsolar abundance \citep[0.49\,Z$_{\sun}$;][]{2010A&A...520A..25Y}.

At a distance of 2300$^{+330}_{-270}$\,pc \citep[][]{2016A&A...595A...1G,2018arXiv180409365G}, V2731\,Oph is the most hard X-ray luminous IP known \citep[log L$_{(X;14-195\,keV)}$\,$>$\,33.9 for $d$\,$>$\,1\,kpc;][]{2014MNRAS.442.2580P}. 
First investigated in X-rays by \citet{2008A&A...481..149D} using XMM-{\it Newton} and INTEGRAL observations, V2731\,Oph was revealed to have a complex X-ray emission. The investigation in a broadband energy (0.2-100\,keV) showed two optically thin plasma components, one cold ($k$T\,$\sim$\,0.2\,keV) and another hot ($k$T\,$\sim$\,60\,keV), with an additional contribution of a hot blackbody ($k$T\,$\sim$\,90\,eV). The intrinsic X-rays suffer the effects of a complex (local) distribution of cold intervening material plus a warm absorber, the latter being suggested by an OVII absorption edge at 0.74\,keV. As with TV\,Col, the pulse modulation for V2731\,Oph depends on the energy. In the case of V2731\,Oph, the X-rays are spin modulated (128.02$\pm$0.02\,s), 
while UV emission is unmodulated at the 3$\sigma$ level \citep{2008A&A...481..149D}.

More recently, both TV\,Col and V2731\,Oph were observed with NuSTAR and investigated in the 3-50\,keV band by \citet{2016ApJ...826..160H} in the context of the X-ray emission of the Galactic center. We revisit those observations for a more in-depth investigation of the X-ray reflection and absorption in these systems, and consequently the characterization of IPs in general. We include in the investigation the Swift/XRT observation carried out during the NuSTAR observation of TV\,Col, and we show that the inclusion of soft X-rays is crucial to disentangle the effects of absorption and reflection in IP systems. For the same reason, we discuss the constraints imposed by previous XMM-{\it Newton}, Suzaku, and Swift observations of V2731\,Oph. Also, contrary to the previous study with NuSTAR data \citep{2016ApJ...826..160H}, we fit the spectra using the cooling flow model that takes into account the ionized Fe\,K lines.

\section{Observations and data reduction}
\label{sct:obs}

Details of the X-ray observations used in this investigation are given in Table \ref{tbl:obs}. 
TV\,Col was observed by the NuSTAR satellite on 2014-05-11 for 49.7 ks (ObsID 30001020002). The system was also simultaneously observed by the Neil Gehrels Swift Observatory (hereafter Swift) for $\sim$\,1.9\,ks on 2014-05-12 (ObsID 00080734001). For the first time, we explored the corresponding NuSTAR FPMA/FPMB and Swift/XRT datasets to carry out consistent wide-band spectroscopy from 0.35 to 75 keV. We also investigate five other Swift/XRT observations in order to obtain clues on the spectral variations of the source.

We investigated V2731\,Oph using a NuSTAR observation carried out for $\sim$\,50\,ks on 2014-05-14 (ObsID 30001019002), for which there is no contemporaneous X-ray observation with another satellite. However, we used archival observations, namely XMM-{\it Newton}/EPIC (2005-08-29, by 13.5\,ks; ObsID 0302100201), Suzaku (2009-02-16, by 32.9\,ks; ObsID 403026010), and the longest Swift (2007-02-23, by 12.8\,ks; ObsID 00035086002) observations in order to evaluate the source in soft X-rays and discuss implications to the description of the system.

The data reduction followed standard procedures with the appropriated tools (with HEASOFT-v6.22 for the NuSTAR, Swift, and Suzaku data, and SAS-v16.1.0 for the XMM-{\it Newton} data), with the latest calibration files that were available in 2017 November.
We use different extraction regions for the NuSTAR modules, FPMA and FPMB, based on their individual images, to accommodate the relative astrometric offset between them. 

All spectra from both TV\,Col and V2731\,Oph were binned such that each energy bin contains at least 25 counts, except for the Swift/XRT data of TV\,Col. For those data a 5 counts limit was adopted 
 as enforcing 25 counts per bin, making Gaussian errors a
    reasonable approximation, would have meant too few spectral bins
    for the spectrum to be useful.
Spectral fits were done with Xspec v12.9.1m by using the $\chi$-squared as fit and test statistics, but applying the C statistic for the fit in all cases in which the Swift/XRT data of TV\,Col were investigated 
 -- as it is appropriate to low count per energy bin, following the deviation from Gaussian to Poisson data.

\begin{table}
\begin{center}
\caption{Journal of X-ray observations. \label{tbl:obs}}
\begin{tabular}{lcccc}
\tableline\tableline
\tableline
& ObsID & Date & Exposure \\
\tableline
TV\,Col \\
\tableline
Swift		& 00035282001 & 2007-03-30 & 7\,ks    \\
                & 00035282002 & 2007-04-23 & 0.8\,ks  \\
                & 00037150002 & 2007-12-21 & 19.4\,ks \\
                & 00037150003 & 2007-12-27 & 4.6\,ks  \\
                & 00037150005 & 2008-01-14 & 5.3\,ks  \\
                & 00080734001 & 2014-05-12 & 1.9\,ks \\
NuSTAR	& 30001020002 & 2014-05-11 & 49.7\,ks \\
\tableline
V2731\,Oph\\
\tableline
XMM-{\it Newton}& 0302100201  & 2005-08-29 & 13.5\,ks\\
Swift		& 00035086002 & 2007-02-23 & 12.8\,ks\\
Suzaku		& 403026010   & 2009-02-16 & 32.9\,ks\\
NuSTAR 	& 30001019002 & 2014-05-14 & 50\,ks \\
\tableline
\end{tabular}
\end{center}
 Notes: The covered energy range was 0.3-10\,keV for the XMM-{\it Newton}, Swift, and Suzaku satelites; for NuSTAR, it was 3--60\,keV for TV\,Col and 3--75\,keV (or 20--75\,keV; see text) for V2731\,Oph.\\
\end{table}

\begin{table*}
\begin{center}
\caption{Spectral models. \label{tbl:models}}
\begin{tabular}{llllllll}
\tableline\tableline
\tableline
Target & Model & Description \\
\tableline

TV\,Col & M1 & \textsc{constant*phabs*[(apec} or \textsc{mkcflow) + gauss]} \\ 
	& M2 & \textsc{constant*phabs*[reflect*(apec} or \textsc{mkcflow) + gauss]} \\
	& M3 & \textsc{constant*phabs*pwab*[(apec} or \textsc{mkcflow) + gauss]}\\
	& M4 & \textsc{constant*phabs*pwab*[reflect*(apec} or \textsc{mkcflow) + gauss]} \\
	& M5 & \textsc{constant*phabs* pwabs*[reflect*(apec} or \textsc{mkcflow) + (apec} or \textsc{bbody) + gauss]} \\

\tableline

V2731\,Oph & M6 & \textsc{constant*phabs*pwab*edge*(apec+mkcflow+bbody+gaussian)} \\
           & M7 & \textsc{constant*phabs*pwab*edge*(apec+reflect*mkcflow+bbody+gaussian)} \\
           & M8 & \textsc{constant*phabs*pwab*edge*(vapec+vmcflow+gaussian)}\\           
           & M9 & \textsc{constant*phabs*pwab*edge*(vapec+vmcflow+bbody+gaussian+gaussian+gaussian)}\\           
           
\tableline
 
\end{tabular}
\end{center}
\end{table*}

\section{X-ray spectroscopy}

Throughout this work, we use \textsc{xspec} and models available within it to carry out modeling of XMM-{\it Newton}, Swift, Suzaku, and NuSTAR X-ray spectra. Two {\it optically thin} thermal models were used -- a single temperature \textsc{apec} model and a multi-temperature cooling flow model with \textsc{mkcflow}, also based on \textsc{apec} (with switch parameter equal to 2), which is model M1. In addition, an optically thick blackbody model, \textsc{bbody}, was included as necessary. The abundance table applied in both optically thin thermal models was that of \citet{2009ARA&A..47..481A}. The redshift required by construction in the \textsc{mkcflow} model cannot be zero. It was assumed to be 8.5867$\times$10$^{-8}$ for TV\,Col 
and 5.367$\times$10$^{-7}$ for V2731\,Oph, 
as estimated from the GAIA DR2 distances \citep[513$^{+4}_{-5}$\,pc and 2300$^{+330}_{-270}$\,pc, respectively;][]{2016A&A...595A...1G,2018arXiv180409365G}  and standard cosmological values of \textsc{xspec}. The low temperature  of the \textsc{mkcflow} was fixed to the minimum value allowed by the model ($k$T\,=\,80.8\,eV). 

The interaction of the radiation with matter was accounted by the following models. 
For the photoelectric absorption by cold matter, we use both the
single absorber \textsc{phabs} and the complex absorber \textsc{pwab} \citep{1998MNRAS.298..737D} components. The usual partial covering
absorber models (either the \textsc{pcfabs} multiplicative model or the
\textsc{partcov} convolution model combined with any absorption model)
represent two lines of sights, one with and one without an intervening
absorber. However, the absorption in the pre-shock accretion columns of
magnetic CVs requires a model with a continuous distribution of
covering fraction as a function of column height. \citet{1998MNRAS.298..737D}
developed \textsc{pwab} to represent this situation in which the covering
fraction is assumed to be a power-law function of the column ($\beta$ index, 
from a minimum (N$_{H,min}$) to a maximum (N$_{H,max}$) equivalent hydrogen columns).
An  \textsc{edge} component was used to account for a warm absorber, only necessary in the case of V2731\,Oph, associated with the OVII absorption edge at 0.74\,keV.
The \textsc{reflect} model was applied when investigating reflection of X-rays in the systems -- in all cases keeping the (unconstrained) inclination angle ($i$) of the reflecting surface equal to the default value in the model, such that cos($i$) is equal to 0.45.

The \textsc{gauss} model was used to describe the 6.4\,keV fluorescence iron line, with the line energy (E$_{c}$) and width ($\sigma$) fixed to 6.4\,keV and 0.01\,keV, respectively.  
An energy-independent multiplicative factor, the \textsc{constant} model, was included to account for the cross-calibration uncertainties of the different instruments (namely, the impact on normalization in spectral modeling) when two or more spectra were used in simultaneous fits. 
\textsc{constant} also helps with source flux variability, since the observations are not all simultaneous.
All parameters cited above as being fixed were not constrained when they were allowed to vary freely during the fit.

We notice that the statistical significance for the NuSTAR and Swift simultaneous fit of the TV\,Col spectra is biased to the NuSTAR data, and thus the visual check was crucial to choose between the models.
For TV\,Col, we started with the simplest model based on single absorption for the two optically thin thermal models cited above (\textsc{apec} and \textsc{mkcflow}; M1),
then adding the reflection model (M2), 
replacing reflection model by the partial-covering model (M3),
and finally using both partial-covering and reflection models (M4).
In all cases, a Gaussian line at 6.4 keV was added to describe the Fe\,K$\alpha$ fluorescence feature. 
Some Swift spectra of TV\,Col presented an excess in soft X-rays which was accounted for as an additional thermal component (M5). Thus, the four initial models are a subgroup of M5 as described in Table \ref{tbl:models}.
For V2731\,Oph, the starting point was the result of \citet{2008A&A...481..149D} from XMM-{\it Newton} and INTEGRAL data. Finally, the model applied to the system was: \textsc{constant*phabs*pwab*edge*(mkcflow+apec+bbody+ gaussian)}.
Table \ref{tbl:models} summarizes the models. 

When using the \textsc{reflect} component we extended the energy range over which the model is calculated to 100\,keV, because photons with energies above the instrumental coverage can be Compton down-scattered to the energy range which is covered by the instrument (command \textsc{energies extend high 100.0}, in \textsc{xspec}). The output spectrum was limited to 80\,keV to speed up the calculation of the spectrum (command \textsc{xset reflect\_max\_e  80.0} in \textsc{xspec}).

\begin{figure*}
\centerline{
\includegraphics[angle=-90,scale=1]{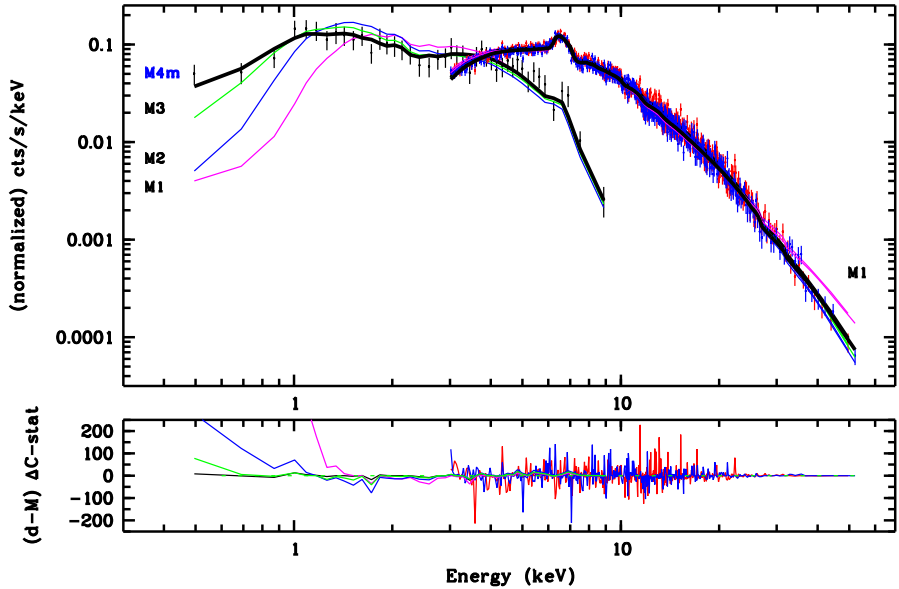}
}
\caption{TV\,Col: NuSTAR/FPMA and Swift/XRT spectra fitted with models M1 (cyan), M2 (blue), M3 (green), and M4$_{\rm m}$ (black; see text for futher details). The lower panel
shows the residual with respect to M4$_{\rm m}$.} \label{fig:tvspct}
\end{figure*}

\subsection{TV\,Col}

TV\,Col is among the brightest hard X-ray IPs \citep{2014MNRAS.442.2580P}. Its X-rays are due to optically-thin, thermal plasma, as supported by the strong Fe\,K line complex. 
Because its Galactic latitude of -30.6 deg and distance of 513$^{+4}_{-5}$\,pc, most of the interstellar medium (ISM) in its direction is between us and TV Col. However, the system occupies a relatively unobscured part of the Galaxy with an integrated HI column density estimate of only $\sim$\,2.39$\times$10$^{20}$\,cm$^{-2}$ \citep{2005A&A...440..775K}. 
Thus, the X-rays from TV\,Col are only slightly affected by the ISM and any absorption we measure is due to local contributions. 
These circumstances make it an ideal target for a detailed study of intrinsic absorbers in the system and therefore of reflection effects. We base our spectral analysis on these two premises and proceed from the simplest model to more complex ones, until satisfactory fits were obtained.

\subsubsection{Broad band spectroscopy from simultaneous Swift and NuSTAR observations}
\label{sct:TVCol_joiningfit}

We first investigate TV\,Col by simultaneously fitting the contemporaneous NuSTAR  (FPMA\,+\,FPMB; 3--60\,keV) and Swift/XRT spectra (0.35--10\,keV; Sect. \ref{sct:obs}) that together covered the wide energy range from 0.35 to 60 keV. The iron line complex (via NuSTAR) and the soft part of the spectrum (via XRT) were always included, which played a key role in disentangling the models.

A thermal model affected by a single absorption component (M1) with either \textsc{apec} and \textsc{mkcflow} does not describe the data. It fails in the description of the continuum, especially at low energies (see Fig. \ref{fig:tvspct}). However, this simple model immediately revealed the presence of a complex absorption pattern in the system.  This model underestimates the flux for E\,$<$\,1.5\,keV appreciably, showing that the absorber with a high column density that is required to explain the spectral shape at higher energies absorbs too many soft photons.
Similar results were obtained when including the reflection component (M2).
The fit was substantially improved when the reflection component was replaced by the partial-covering model (M3), but still yielded an unacceptable description for energies up to $\sim$\,6\,keV, below which the model oscillates between underestimating and overestimating the flux. Finally, the reflection was added to M3, by using the M4 model. This model significantly improved the fit over the whole spectrum, as described below.

\begin{table}
\begin{center}
\caption{TV\,Col: the best-fit spectral parameters from the simultaneous fit of the contemporaneous NuSTAR (FPMA \& FPMB) and Swift/XRT spectra from best fit M4$_{\rm m}$ model (as in Table \ref{tbl:models}). \label{tbl:spct_tvcol}}
\begin{tabular}{llllllll}
\tableline\tableline
\tableline
Component & Parameter  & Value \\
\tableline
\textsc{phabs}   & N$_{H}$ (10$^{22}$\,cm$^{-2}$)      &0.0239$^{(*)}$\\
\tableline 
\textsc{pwab}    & N$_{H,min}$ (10$^{22}$\,cm$^{-2}$)  & 10$^{-7}$\,$^{(*)}$\\
                 & N$_{H,max}$ (10$^{22}$\,cm$^{-2}$)  & 10.5$^{+1.7}_{-0.8}$\\
                 & $\beta$                             &-0.41$^{+0.03}_{-0.04}$\\
\tableline
\textsc{reflect} & rel$_{refl}$                        & 0.88$^{+0.13}_{-0.13}$\\
                 & cos($i$)                            & 0.45\,$^{(*)}$\\
\tableline
\textsc{mkcflow} & kT$_{low}$ (keV)                    & 0.0808\,$^{(*)}$ \\
                 & kT$_{max}$ (keV)                   & 31.0$^{+1.1}_{-1.1}$\\
		 & Z ($\times$Z$_{\odot}$)             & 0.49$^{+0.04}_{-0.04}$\\
		 & redshift                            & 1.197$\times$10$^{-7}$\,$^{(*)}$ \\
                 & switch                              & 2\\
\tableline
$\chi^{2}_{\nu}$/d.o.f. &                              & 0.95/1007 \\		
\tableline
\end{tabular}
\end{center}
Notes: (*) fixed parameter (see text).\\ 

\end{table}

Contrary to the first three models, M4 describes the continuum from 0.35 to 60\,keV (see Fig. \ref{fig:tvspct}).
The distinction between the model M4 using \textsc{apec} (M4$_{\rm a}$) or \textsc{mkcflow} (M4$_{\rm m}$) comes from the description of the Fe K complex and the surrounding continuum. While the spectral fit from \textsc{apec} substantially  underestimates the flux from the ionized Fe\,XXV line, the \textsc{mkcflow} model describes both it and the Fe\,XXVI lines and results in a better description of the continuum at $\sim$\,4--8\,keV -- M4$_{\rm m}$, the assumed model for TV\,Col in this work, resulting in $\chi^2_{\nu}$ of 0.95 for 1007 degrees of freedom.

We present the best-fit parameters for M4$_{\rm m}$ in Table \ref{tbl:spct_tvcol}. The equivalent hydrogen column from the \textsc{phabs} model is constrained to be less than 3.0$\times$10$^{20}$\,cm$^{-2}$ such that any intrinsic absorption would be mitigated by the effect of the dominant local partial-covering absorber. Thus, although the values of all parameters are still mutually consistent at 1$\sigma$, we assume a conservative upper limit by fixing the column value to the total Galactic column in the direction of TV\,Col \citep[2.39$\times$10$^{20}$\,cm$^{-2}$;][]{2005A&A...440..775K}. 
This procedure improves the determination of the absorption derived from the partial-covering absorption \textsc{pwab} model, resulting in N$_{H,max}$\,=\,1.05$^{+0.17}_{-0.08}$$\times$10$^{23}$\,cm$^{-2}$ and $\beta$\,=\,$-$0.41$^{+0.03}_{-0.04}$. The primary source of X-rays is a high-temperature plasma with an abundance of 0.49$\pm$0.04 times the solar values, cooling down from a maximum temperature $k$T of 31.0$\pm$1.1\,keV. If we take this as the measurement of the gravitational potential at the surface of the WD, then its mass is 0.735$\pm$0.015 M$_\odot$ \citep[following][]{1973PThPh..49.1184A}. Our subsequent analysis justifies this assumption, as we discuss in Section \ref{sct:tandrefl}.

We find that the reflection of the primary X-rays makes a non-negligible contribution to the observed X-rays. According to the best-fit model (M4$_{\rm m}$), the reflection amplitude is constrained to be 0.88$^{+0.13}_{-0.13}$. 
The reflection accounts for about 16\% out of a total unabsorbed flux of $\sim$\,1.37$\times$10$^{-10}$ erg\,s$^{-1}$\,cm$^{-2}$ at 0.3--75\,keV, increasing to 30\% out of a total of $\sim$\,5.3$\times$10$^{-11}$ erg\,s$^{-1}$\,cm$^{-2}$ at 10--50\,keV (and 7.6\% out of $\sim$\,7.9$\times$10$^{-11}$ erg\,s$^{-1}$\,cm$^{-2}$ at 0.3--10\,keV). TV\,Col was observed with a luminosity of 4.3$\times$10$^{33}$\,erg\,s$^{-1}$ at 0.3--75\,keV on 2014-05-11/12, and a bolometric X-ray luminosity of (4.6$\pm$0.3)$\times$10$^{33}$\,erg\,s$^{-1}$.

The NuSTAR data suggest a pronounced fluorescent iron line at 6.4\,keV with an equivalent width (EW) estimated in 179$^{+19}_{-61}$ eV. As we show below, this value is significantly higher than that derived for V2731\,Oph with the same instrument, even though V2731\,Oph is significantly more absorbed than TV\,Col. Thus, the fluorescent line at least for TV\,Col seems to have the contribution not only from the absorber but also from the reflector.

\subsubsection{Spectral evolution from Swift/XRT observations}

We investigated individually five other archival Swift/XRT observations of TV\,Col in order to evaluate its spectral evolution (Table \ref{tbl:obs}). Thermal models resulted in unconstrained values for the plasma temperature and abundance while they are consistent with those obtained from the simultaneous NuSTAR and Swift observations. Subsequently, we fitted the spectra by assuming the model M4$_{\rm m}$, keeping fixed the plasma temperature and abundance values of the \textsc{mkcflow} and the parameters of the \textsc{reflect} component to those determined previously from the NuSTAR/Swift observations of 2014-05-11/12, presented in Table \ref{tbl:spct_tvcol}.
We also adopted the same fixed parameter values as done previously for the M4$_{\rm m}$ model. The best-fit spectral parameters are summarized in Table \ref{tbl:tvcolspctparam}.

There is an excess emission below $\sim$\,1\,keV that indicates the presence of an additional thermal component for two observations (0003715002 and 0003715005), and which is suspected for two others (00035282001 and 0003715003). Its description suggests that it is more likely due to a cool optically thin plasma ($k$T\,$\sim$\,0.15\,keV) but a blackbody emission ($k$T\,$\sim$\,0.12\,keV) cannot be ruled out. 
No strong change in spectral shape of TV\,Col is observed in the first five Swift observations listed in Table \ref{tbl:obs} but they still exhibit variability in the absorbers and also in the unabsorbed flux at 0.3-10\,keV. Remarkably, during the fifth observation, the values increased by 50--100\%.

\begin{table*}
\begin{center}
\caption{TV\,Col: best-fit spectral parameters from Swift/XRT observations. \label{tbl:tvcolspctparam}}
\begin{tabular}{cccccccc}
\tableline\tableline
\tableline
		& \multicolumn{2}{c}{\textsc{pwab}}     	 & \textsc{apec}          & \textsc{bbody}              &                       &                         \\
\cline{2-3}
Swift/XRT	& N$_{H,max}$          &  $\beta$                & $k$T    	          & $k$T                        & F$_{unabs,0.3-10keV}$ & $\chi^{2}_{\nu}$/d.o.f. \\
(ObsID)		& (10$^{22}$\,cm$^{-2}$)&                        & (keV)                  & (keV)                       & ($\times$10$^{-11}$\,erg\,cm$^{-2}$\,s$^{-1}$)& \\
\tableline
00035282001 & 13.8$^{+3.4}_{-2.1}$  & -0.41$^{+0.03}_{-0.03}$ & ...                    & ...                         & 6.5                   & 1.04/398                \\
            & 12.7$^{+2.6}_{-2.7}$  & -0.38$^{+0.04}_{-0.01}$ & 0.14$^{+0.06}_{-0.04}$ & ...                         & 6.7                   & 1.04/396                \\
            & 12.7$^{+3.0}_{-3.2}$  & -0.38$^{+0.09}_{-0.02}$ & ...                    & 0.11$^{+0.06}_{-0.05}$      & 6.6                   & 1.04/396                 \\
00035282002 & 13.6$^{+8.0}_{-5.1}$  & -0.40$^{+0.09}_{-0.06}$ & ...                    & ...                         & 6.9                   & 0.88/65                 \\
00037150002 & 17.2$^{+3.9}_{-1.4}$  & -0.37$^{+0.02}_{-0.03}$ & 0.18$^{+0.01}_{-0.02}$ & ...                         & 8.6                   & 1.12/689                \\
            & 15.9$^{+2.1}_{-0.2}$  & -0.35$^{+0.01}_{-0.02}$ & ...                    & 0.12$^{+0.01}_{-0.01}$      & 8.5                   & 1.13/689                \\
00037150003 & 13.4$^{+3.5}_{-2.9}$  & -0.35$^{+0.06}_{-0.04}$ & 0.18$^{+0.05}_{-0.06}$ & ...                         & 7.8                   & 1.01/297                 \\
            & 10.1$^{+4.2}_{-2.4}$  & -0.23$^{+0.20}_{-0.11}$ & ...                    & 0.15$^{+0.02}_{-0.03}$      & 7.9                   & 1.01/297                 \\
00037150005 & 25.3$^{+17.6}_{-5.4}$ & -0.43$^{+0.02}_{-0.04}$ & 0.15$^{+0.03}_{-0.02}$ & ...                         & 11.2                  & 1.00/389                 \\
            & 31.7$^{+13.6}_{-9.2}$ & -0.45$^{+0.03}_{-0.03}$ & ...                    & 0.09$^{+0.02}_{-0.02}$      & 11.6                  & 1.03/389                 \\
\tableline

\end{tabular}
\end{center}
Notes: from M4$_{\rm m}$ or M5 (see Table \ref{tbl:models}).\\
\end{table*}

\subsection{V2731\,Oph}

It is clear from the investigation of TV\,Col that a simultaneous or at least contemporaneous broadband coverage is essential in describing the effects of both absorption and reflection of X-rays and hence in characterizing the system. Even though V2731\,Oph has been observed with high signal-to-noise ratio in soft and hard X-rays, the coverage in both energy bands with different satelites was not contemporaneous. 

\begin{figure*}
\centerline{
\includegraphics[angle=-90,scale=1]{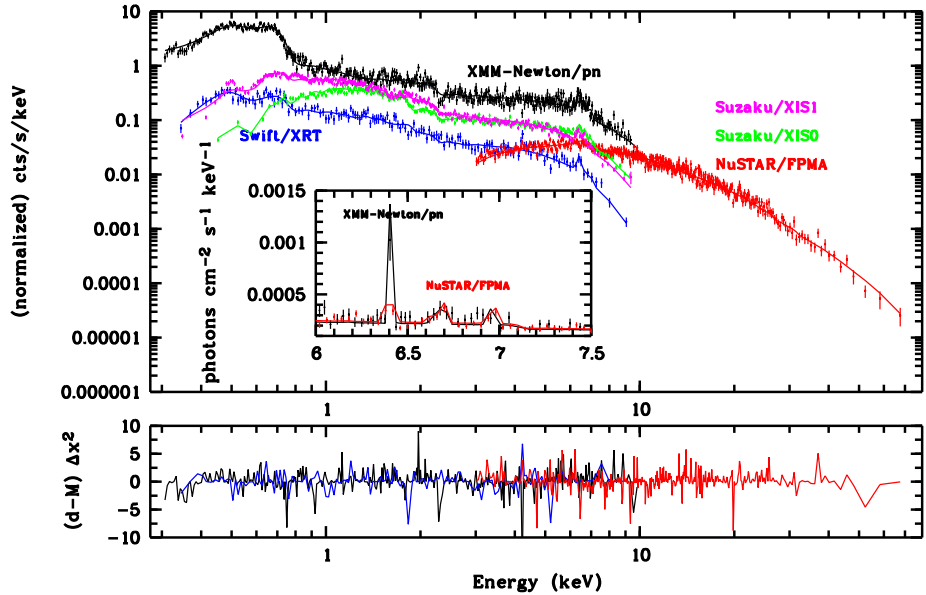}
}
\caption{V2731\,Oph: XMM/pn, Swift/XRT, NuSTAR/FPMA, Suzaku/XIS0, and Suzaku/XIS1 spectra and the corresponding descriptions from M6. The inset presents the unfolded XMM/pn and NuSTAR/FPMA data and modeling around the Fe\,K complex. For Suzaku/XIS0 and XIS1, see zoom in Fig. \ref{fig:v2731Suzaku}. \label{fig:v2731ophspct}}
\end{figure*}

We started by comparing the previous XMM-{\it Newton} observation (carried out on 2005-08-29) already reported by \citet{2008A&A...481..149D} with the Suzaku \citep[2009-02-16;][]{2010A&A...520A..25Y} and the longest Swift (2007-02-23) archival observations of V2731\,Oph (Table \ref{tbl:obs}). 
Those observations were investigated separately and, subsequently, via a joint fit with the 20-75 keV NuSTAR (2014-05-14) spectra. 
For the purpose of obtaining the based fit of the XMM-{\it Newton}
    data below 10 keV, we fit them jointly with the NuSTAR spectra
    restricted to 20-75 keV range, allowing us to utilize the
    parameters that best describe the high energy shape of the
    spectrum. In this procedure, we ignored the NuSTAR data below
    20 keV, because of our suspicion (later confirmed) that there
    is more spectral variability at low energies.
We then proceeded with an investigation of the NuSTAR observations of the system separately using the whole available spectral range (3-75\,keV), and then with the Swift and Suzaku spectra, also separately, in all cases fixing the values of the cooling flow component to those derived from the simultaneous fit of the XMM-{\it Newton} and the 20-75 keV NuSTAR spectra. Figure \ref{fig:v2731ophspct} shows the spectra.

\subsubsection{Clues about the spectral evolution}
\label{sect:V2731_spctevol}

As reported by \citet{2008A&A...481..149D}, an XMM-{\it Newton} observation revealed a complex X-ray spectrum of V2731\,Oph.
A good fit required multiple thermal plasma components, with both a cold and another hot optically thin plasmas, and an additional blackbody emission. Once produced, the X-rays are affected by a complex distribution of cold matter and also suffers absorption from a warm material, as indicated by the OVII absorption edge at 0.74\,keV. The authors showed that the spectral distribution is well explained using the model \textsc{wabs*pcfabs*edge(mekal+bbody+mekal+gaussian)}.

We first applied the model used by \citet{2008A&A...481..149D} to carry out a simultaneous fit of the three EPIC XMM-{\it Newton} spectra (pn, MOS1, and MOS2) and reproduced their results. However, in the following we adopt the (updated) \textsc{phabs} model instead of \textsc{wabs}, and the (more appropriate) \textsc{pwab} model in the place of \textsc{pcfabs}. Also, we replace the \textsc{apec} describing the hottest thermal component by a \textsc{mkcflow} model, in line with the case of our other target, TV\,Col, and as it is expected for IPs. Thus, the final model is M6:  \textsc{constant*phabs*pwab*edge*(mkcflow+apec+bbody+ gaussian)}.

The model M6 describes the XMM-{\it Newton} data well, resulting in $\chi^{2}_{\nu}$\,=\,1.05 for 834 degrees of freedom, but the hottest component of the \textsc{mkcflow} component converged to the hard limit of $k$T\,=\,79.9\,keV. The coldest component, as for the case of TV\,Col, was not constrained and therefore fixed to $k$T\,=\,80.8\,eV. We added to the fit the NuSTAR spectra limited to the 20--75\,keV energy range. This inclusion improves the statistic to $\chi^{2}_{\nu}$\,=\,1.01, for 962 degrees of freedom, while resulting in well-constrained values (Table \ref{tbl:v2731spctparam}). Starting with the plasma parameters, the temperature derived from the \textsc{mkcflow} model peaks at $k$T\,=\,47.6$^{+4.2}_{-4.5}$\,keV.
A cold ($k$T\,=\,0.17$^{+0.01}_{-0.02}$\,keV), optically thin plasma component is necessary to account for a substantial part of the $\sim$\,0.5--1.5\,keV range which is not described by the \textsc{mkcflow} model even taking its low temperature as a free parameter during the fit. A fit without the cold plasma, but with the low temperature left to vary free, only achieves $\chi^{2}_{\nu}$\,=\,1.15.
On the other hand, we confirm that a blackbody component (with $k$T\,=\,0.094$^{+0.002}_{-0.002}$\,keV) is a good description for the excess emission in soft X-rays, not explained by the optically thin plasma emission. However, this is unphysical for a white dwarf, as it is argued at the end of this section. In terms of absorption, the modeling with \textsc{pwab} resulted in a complex structure with N$_{H,max}$\,=\,156$^{+270}_{-42}$$\times$10$^{22}$\,cm$^{-2}$ and $\beta$\,=\,-0.68$^{+0.01}_{-0.01}$. Although poorly constrained, the lower limit to N$_{H,max}$ is still high: at least $\sim$\,10$^{24}$\,cm$^{-2}$. The N$_{H,min}$ parameter, unconstrained, was fixed to 10$^{15}$\,cm$^{-2}$ -- the minimum value allowed by the model. The \textsc{phabs} resulted in the equivalent to N$_{H}$\,=\,(3.3$\pm$0.2)$\times$10$^{21}$\,cm$^{-2}$. 
 Such a value is higher than the total Galactic HI column density in the direction of V2731\,Oph of about 1.6$\times 10^{21}$ cm$^{-2}$ estimated by  \citet{2005A&A...440..775K} but slightly lower than the value of 3.9$\times 10^{21}$ cm$^{-2}$ as derived from the 3D Dust Mapping with Pan-STARRS 1\footnote{http://argonaut.skymaps.info/} \citep{2018MNRAS.478..651G}.
The abundance was tied to the \textsc{apec} and \textsc{mkcflow} models, resulting in Z\,=\,0.30$\pm$0.09\,Z$_{\odot}$. This value is consistent with the two values reported by \citet{2008A&A...481..149D}, of 0.33$^{+0.37}_{-0.19}$\,Z$_{\odot}$ from the 0.3-10\,keV XMM-{\it Newton} EPIC data alone and 0.40$^{+0.07}_{-0.08}$Z$_{\odot}$ when the 20-100 keV INTEGRAL/ISGRI spectrum was also included in the analysis. The energy of the \textsc{edge} component was fixed at 0.74\,keV \citep[corresponding to the OVII absorption edge reported by][]{2008A&A...481..149D}.

The aim of this analysis is to compare how the spectral energy distribution has evolved when comparing observations carried out at different epochs. In this exercise we explore the longest archival Swift observations and the Suzaku observation of V2731\,Oph, verifying the consistency of the spectra with the model used to describe the 0.3--10\,keV XMM-{\it Newton} plus the 20--75\,keV NuSTAR  spectra. 

We first investigate the Swift/XRT spectrum (ObsID 00035086002). 
The model and values derived from the joint XMM-{\it Newton} and NuSTAR analysis, keeping only the normalization as free parameters, results in an unacceptable description of the Swift spectrum ($\chi^{2}_{\nu}$\,=\,3.18). 
The statistic is slightly improved when allowing the parameter of the \textsc{phabs} to vary ($\chi^{2}_{\nu}$\,=\,2.54). 
No improvement is obtained by assuming the N$_{H,max}$ parameter of the \textsc{pwab} free during the fit, while it is not constrained, but a significant improvement is obtained allowing the $\beta$ to vary ($\chi^{2}_{\nu}$\,=\,1.59) -- keeping the lower limit for the absorption still fixed. 
The statistic improves to $\chi^{2}_{\nu}$\,=\,1.14 when the absorption depth of the \textsc{edge} component is also a free parameter. 
Allowing the plasma temperature of the \textsc{apec} component to also vary improves the fit description ($\chi^{2}_{\nu}$\,=\,1.06). 
On the other hand, there is no improvement when allowing the temperature 
of the \textsc{bbody} component and the high temperature parameter of the \textsc{mkcflow} model to vary. 
As their values are consistent with those derived from the XMM-{\it Newton} plus NuSTAR analysis, we kept them fixed to the corresponding values. The best-fit spectral parameters are summarized in Table \ref{tbl:v2731spctparam}. The analysis revealed a significant change in the partial-covering absorption and in the temperature of the cold plasma component, while the spectral energy distribution of hard X-rays remained essentially unchanged.
\begin{table*}
\begin{center}
\caption{V2731\,Oph: best-fit spectral parameters. \label{tbl:v2731spctparam}}
\begin{tabular}{ccccccccccc}
\tableline\tableline
\tableline
		&\textsc{phabs}& \multicolumn{2}{c}{\textsc{pwab}$^{1}$}     	& \textsc{edge}$^2$ & \textsc{apec}          & \textsc{bbody}              & \textsc{mkcflow}$^{3}$ &                       &                         \\
		\cline{3-4}   
         	&N$_{H}$   & N$_{H,max}$          &  $\beta$             &  abs. depth & $k$T    	          & $k$T                        & $k$T$_{max}$   &$Z$$^4$ &   $\chi^{2}_{\nu}$/d.o.f. \\

         	\cline{2-3} \cline{6-8}
         	& \multicolumn{2}{c}{(10$^{22}$\,cm$^{-2}$)}         && ($@$0.74\,keV) & \multicolumn{3}{c}{(keV)}                       & ($Z_{\odot}$)    & \\
\tableline
\tableline
\tableline
\multicolumn{5}{l}{Without \textsc{reflect} from M6 and its variants:}\\
\tableline
\multicolumn{8}{l}{Model: \textsc{constant*phabs*pwab*edge*(apec + mkcflow + bbody + gaussian)}} \\
XMM\,+\,NuSTAR$^5$ & 0.33$^{+0.02}_{-0.02}$  & 156$^{+270}_{-42}$ & -0.68$^{+0.01}_{-0.01}$ & 1.86$^{+0.08}_{-0.09}$ & 0.17$^{+0.01}_{-0.02}$ & 0.094$^{+0.002}_{-0.002}$ & 47.6$^{+4.2}_{-4.5}$ & 0.30$^{+0.09}_{-0.09}$ & 1.01/962                \\
Swift & 0.29$^{+0.03}_{-0.03}$ & 156$^6$ & -0.75$^{+0.04}_{-0.04}$ & 1.46$^{+0.22}_{-0.24}$ & 0.64$^{+0.05}_{-0.06}$ & 0.094$^2$ & 47.6$^6$ & 0.30$^6$  & 1.02/150 \\
\tableline
\multicolumn{8}{l}{Model: \textsc{constant*pwab*edge*(apec + mkcflow + bbody + gaussian)}} \\

Suzaku (XIS0+3) & ... & 156$^6$ & -0.78$^{+0.01}_{-0.01}$ & 0.72$^{+0.07}_{-0.07}$ &  0.25$^{+0.01}_{-0.01}$ & 0.21$^{+0.01}_{-0.01}$ & 47.6$^6$ & 0.30$^6$  & 1.06/1663 \\

Suzaku (XIS0+1+3) & ... & 156$^6$ & -0.80$^{+0.01}_{-0.01}$ & 0.75$^{+0.04}_{-0.04}$ & 0.28$^{+0.01}_{-0.01}$ & 0.18$^{+0.01}_{-0.01}$ & 47.6$^6$ & 0.30$^6$  & 1.18/2574 \\

\tableline
\multicolumn{8}{l}{Model: \textsc{constant*pwab*(mkcflow + gaussian)}} \\
NuSTAR (3-75\,keV) & ... & 753$^{+60}_{-62}$ & -0.68$^{+0.01}_{-0.01}$ & ... & ... & ... & 47.6$^6$ & 0.30$^6$ & 1.00/782\\ 
NuSTAR (3-75\,keV) & ... & 766$^{+131}_{-88}$ & -0.68$^{+0.01}_{-0.01}$ & ... & ... & ... & 46.7$^{+3.6}_{-3.7}$ & 0.30$^6$ & 1.02/781\\ 
\tableline
\tableline
\tableline
\multicolumn{5}{l}{With \textsc{reflect} from M7 and its variants:}\\
\tableline
\multicolumn{8}{l}{Model: \textsc{constant*phabs*pwab*edge*(apec + reflect*mkcflow + bbody + gaussian)}} \\
XMM\,+\,NuSTAR$^5$ & 0.36$^{+0.02}_{-0.02}$ & 150$^{+487}_{-62}$ & -0.76$^{+0.01}_{-0.01}$ & 1.83$^{+0.09}_{-0.09}$ & 0.17$^{+0.01}_{-0.02}$ & 0.093$^{+0.002}_{-0.002}$ & 54.5$^{+6.4}_{-4.3}$ & 0.23$^{+0.07}_{-0.05}$  & 0.98/962\\
XMM\,+\,NuSTAR$^5$ & 0.36$^{+0.02}_{-0.02}$ & 121$^{+135}_{-35}$ & -0.76$^{+0.01}_{-0.01}$ & 1.83$^{+0.09}_{-0.09}$ & 0.17$^{+0.01}_{-0.02}$ & 0.092$^{+0.002}_{-0.002}$ & 55.1$^{+6.2}_{-4.4}$ & 0.30$^6$  & 0.98/963\\
Swift & 0.59$^{+0.09}_{-0.11}$ & 121$^6$ & -0.93$^{+0.03}_{-0.03}$ & 1.28$^{+0.24}_{-0.27}$ & 0.22$^{+0.02}_{-0.01}$ & 0.04$^{+0.01}_{-0.01}$ & 55.1$^6$ & 0.30$^6$ &  0.97/149 
\\
\tableline
\multicolumn{8}{l}{Model: \textsc{constant*pwab*edge*(apec + reflect*mkcflow + bbody + gaussian)}} \\

Suzaku (XIS0+3)& ... & 121$^6$ & -0.88$^{+0.01}_{-0.01}$ & 0.74$^{+0.06}_{-0.07}$ & 0.25$^{+0.01}_{-0.01}$ & 0.20$^{+0.01}_{-0.01}$ & 55.1$^6$ & 0.30$^6$  & 1.05/1663 \\

Suzaku (XIS0+1+3) & ... & 121$^6$ & -0.90$^{+0.01}_{-0.01}$ & 0.74$^{+0.04}_{-0.04}$ & 0.30$^{+0.01}_{-0.01}$ & 0.18$^{+0.01}_{-0.01}$ & 55.1$^6$ & 0.30$^6$  & 1.18/2574 \\
\tableline
\multicolumn{8}{l}{Model: \textsc{constant*pwab*(reflect*mkcflow + gaussian)}} \\
NuSTAR (3-75\,keV) & ... & 848$^{+174}_{-167}$ & -0.82$^{+0.01}_{-0.01}$ & ... & ... & ... & 55.1$^6$ & 0.30$^6$ & 0.96/782\\ 
NuSTAR (3-75\,keV) & ... & 895$^{+206}_{-198}$ & -0.82$^{+0.01}_{-0.01}$ & ... & ... & ... & 55.8$^{+5.3}_{-3.8}$ & 0.30$^6$ & 0.98/781\\ 
\tableline

\end{tabular}
\end{center}
Notes: 
$^1$ N$_{H,min}$\,=\,10$^{15}$\,cm$^{-2}$; 
$^2$ threshold energy equal to 0.74\,keV; 
$^3$ $k$T$_{low}$\,=\,80.8\,eV; 
$^4$ $Z$\, tied to the \textsc{apec}, \textsc{mkcflow}, and, when present, \textsc{reflect} components, for all the elements; 
$^5$ using only the 20-75\,keV NuSTAR spectra (FPMA and FPMB); 
$^6$ fixed to the values of the joining XMM\,+\,NuSTAR analysis.
For the \textsc{reflect} component, when applied, the reflection scaling factor is equal to 1 and cos($i$)\,=\,0.45. 
The lack of coverage for E\,$<$\,3\,keV from NuSTAR data does not allow a reliable determination of the flux at 0.3-10\,keV.
For the Suzaku analysis, see text.
\\

\end{table*}

We extended the same methodology to the Suzaku XIS0, XIS1, and XIS3 spectra. Allowing only the normalization to vary during the fit resulted in $\chi^{2}_{\nu}$\,=\,6.30. As for the XRT spectrum, the description is improved when allowing the absorption column of the \textsc{phabs} ($\chi^{2}_{\nu}$\,=\,3.70), the $\beta$ parameter of the \textsc{pwab} model ($\chi^{2}_{\nu}$\,=\,2.34), the absorption depth of the \textsc{edge} component ($\chi^{2}_{\nu}$\,=\,1.54), and the plasma temperature of the \textsc{apec} component  ($\chi^{2}_{\nu}$\,=\,1.42) to vary. The column derived from the \textsc{phabs} model is virtually equal to zero (N$_{H}$\,$\sim$\,10$^{16}$\,cm$^{-2}$) and unphysical, and this component was removed without impacting the fit. The lower limit of the absorption in the \textsc{phabs} converges to the minimum value when assumed as a free parameter. In contrast with the Swift spectrum, the fit for the Suzaku spectra is statistically improved when the temperature of the \textsc{bbody} component is free during the fit ($k$T\,$\sim$\,0.18\,keV; $\chi^{2}_{\nu}$\,=\,1.22). Table \ref{tbl:v2731spctparam} includes the best-fit parameters of the Suzaku observation.

\begin{figure*}
\centerline{
\includegraphics[angle=-90,scale=1]{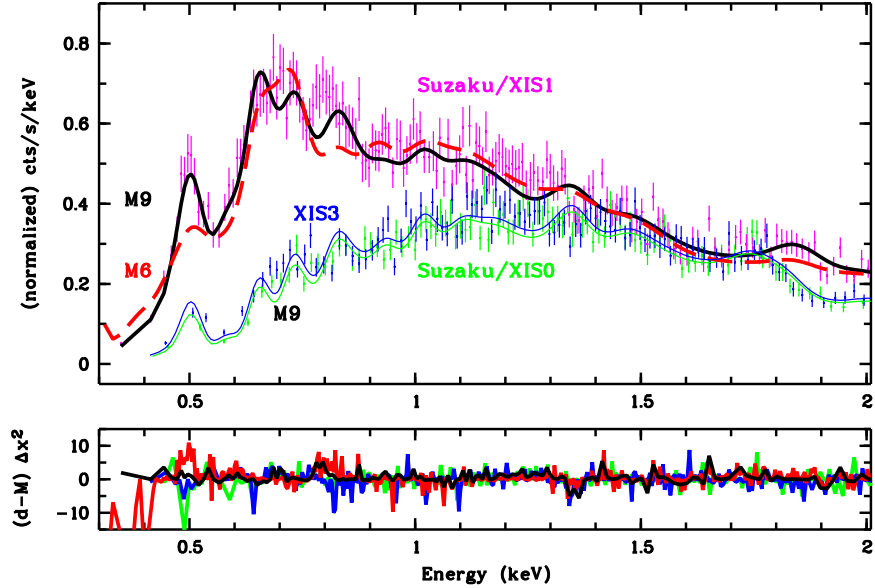}
}
\caption{V2731\,Oph: Suzaku/XIS0 and XIS3 fitted with model M9 (green and blue, respectively) and Suzaku/XIS1 (cyan) with models M6 (red) and M9 (black). See text for futher details.}
\label{fig:v2731Suzaku}
\end{figure*}

However, there are four significant issues with the fit to the Suzaku data. 
First, the model underestimates the flux around 0.5\,keV and also around 0.8 keV (see Fig. \ref{fig:v2731Suzaku}).
Second, the column density for the \textsc{phabs} component is unrealistically low in comparison with the total Galactic HI column density in this direction ($\sim$\,1.6$\times 10^{21}$ cm$^{-2}$ according to \citet{2005A&A...440..775K} and 3.9$\times 10^{21}$ cm$^{-2}$ from the 3D Dust Mapping with Pan-STARRS 1, by  \citet{2018MNRAS.478..651G}). Given the large distance to V2731 Oph (at least 2 kpc), we expect that a substantial fraction of this column is between V2731 Oph and the solar system. The fact that N$_H$ for the \textsc{phabs} component was negligible when
fitting the Suzaku data, while it was higher than the Galactic value in
the XMM-Newton and Swift data, points to a possible problem with the fit.
Third, the blackbody temperature, already worryingly high in the XMM-{\it Newton} fit, is so high to be clearly unphysical for a white dwarf \citep{1987MNRAS.226..725W}; we will return to this point in Section \ref{sct:v2731complex}. 
Fourth, this model does not fit the XIS1 data, which show significant residuals below $\sim$0.8\,keV, especially around 0.5\,keV. While some cross-calibration issues among the XIS units of Suzaku are commonly seen, what we see in V2731\,Oph is far beyond residuals seen in other Suzaku observations. One likely explanation is that the model is wrong but nevertheless works well enough, phenomenologically, with XMM-{\it Newton}, Swift/XRT and Suzaku XIS0+XIS3 data (Table \ref{tbl:v2731spctparam}). In fact, the best-fit spectral parameters from the XIS0+XIS3 data with M6 and M7-like models are similar to those obtained from the XIS0+XIS1+XIS3 data.
Because the Suzaku/XIS1, using an advanced backside-illuminated (BI) CCD chip, has a relatively high effective area below 0.8\,keV and a superior spectral resolution at these energies than those of the other cameras used in this work, some inadequacies of these models becomes readily visible in the XIS1 data. Thus, we explore in the next section an alternative spectral description to V2731\,Oph.

\begin{table*}
\begin{center}
\caption{V2731\,Oph: best-fit spectral parameters by using \textsc{vapec} and \textsc{vmcflow}. \label{tbl:v2731highnitrogen}}
\begin{tabular}{llcccccc}
\tableline
&                                    & M8                      & M8                       & M9\\
&                                    & XMM\,+\,NuSTAR$^1$     & Suzaku                   & Suzaku \\
& & & (XIS0+1+3) & (XIS0+1+3) \\
\tableline
\textsc{phabs} \\
		&N$_{H}$ (10$^{22}$\,cm$^{-2}$)      & 0.14$^{+0.02}_{-0.02}$  & $<$\,0.004               & 0.29$^{+0.02}_{-0.03}$\\
\tableline
\textsc{pwab}$^2$ \\
		& N$_{H,max}$ (10$^{22}$\,cm$^{-2}$) & 612$^{+183}_{-155}$     & 374                      & 68$^{+15}_{-20}$\\
		& $\beta$                            & -0.63$^{+0.01}_{-0.01}$ & -0.79$^{+0.01}_{-0.01}$  & -0.77$^{+0.02}_{-0.01}$\\
\tableline
\textsc{edge}$^3$ \\
		&  Abs. depth ($@$0.74\,keV)         & 1.49$^{+0.08}_{-0.08}$  & 0.52$^{+0.06}_{-0.05}$   & 0.97$^{+0.08}_{-0.06}$\\
\tableline
\textsc{vapec} \\         
		& $k$T (keV)                         & 0.23$^{+0.01}_{-0.01}$  & 0.50$^{+0.01}_{-0.02}$   & 0.49$^{+0.02}_{-0.02}$\\
\tableline
\textsc{vmcflow}$^4$ \\
		& $k$T$_{max}$ (keV)                 & 43.4$^{+5.0}_{-4.0}$    & 43.4$^6$                 & 47.6$^7$\\
\tableline
		& $Z_{N}$ ($Z_{N,\odot}$)$^5$            & 5.8$^{+1.0}_{-0.7}$     & 16.9$^{+2.3}_{-2.1}$     & 69$^{+16}_{-10}$\\
		& $Z_{O}$ ($Z_{O,\odot}$)$^5$            & 0.5$^{+0.1}_{-0.1}$     & 1.3$^{+0.2}_{-0.1}$      & 4.6$^{+0.8}_{-0.5}$\\
		& $Z_{F;Ni}$ ($Z_{F;Ni;\odot}$)$^5$      & 0.37$^{+0.04}_{-0.04}$  & 0.19$^{+0.01}_{-0.02}$   & 0.6$^{+0.1}_{-0.1}$\\
\tableline
\textsc{bbody} \\
		& $k$T (eV)                         & ...                     & ...                      & 39$^{+5}_{-4}$\\
\tableline
\textsc{gaussian}\\
		& Line energy (keV)                       & ...                     & ...                      & 0.58$^{+0.01}_{-0.01}$\\
		& Line energy (keV)                       & ...                     & ...                      & 0.94$^{+0.01}_{-0.01}$\\

\tableline		
		& $\chi^{2}_{\nu}$/d.o.f.            & 1.08/962                & 1.19/2571                & 1.12/2565\\
		
\tableline
\tableline

\end{tabular}
\end{center}
Notes: $^1$ Using only the 20-75\,keV NuSTAR spectra (FPMA and FPMB); 
$^2$ N$_{H,min}$\,=\,10$^{15}$\,cm$^{-2}$; 
$^3$ threshold energy equal to 0.74\,keV; 
$^4$ $k$T$_{low}$\,=\,80.8\,eV; 
$^5$ $Z$\, tied to the \textsc{vapec} and \textsc{vmcflow} components, and fixed to 0.3\,Z$_{\odot}$ for all other elemental abundance; 
$^6$ fixed to the value of the joining XMM\,+\,NuSTAR analysis from M8; $^7$ fixed to the value of the joining XMM\,+\,NuSTAR analysis from M6 (in Table \ref{tbl:v2731spctparam}).
\end{table*}

\subsubsection{Alternative spectral description}

The Suzaku/XIS1 spectrum indicates a more pronounced excess around 0.5\,keV than in other spectra, suggesting the presence of an intense N{\small VII}\,L$\alpha$ line (see red line from M6, in Fig. \ref{fig:v2731Suzaku}). In order to check for an overabundance of N, we replace  \textsc{apec} by \textsc{vapec} and \textsc{mkcflow} by \textsc{vmcflow} and apply the model without reflection to the Suzaku XIS0, XIS1, and XIS3 spectra, simultaneously -- as the variants of the first models that allow the user to set the abundance of the main elements individually. The abundance for the \textsc{vapec} and \textsc{vmcflow} were tied together, allowing Z$_{N}$ to vary freely during the fit but keeping all other elemental abundances equal to 0.30\,Z$_{\sun}$. The fit improves significantly with a better description of the soft tail ($\chi^2_{red}$ from 1.50 to 1.25; yet unnaceptable and not presented). It indicates  Z$_{N}$\,=\,3.2($\pm$0.3)\,Z$_{\odot}$ while the normalization of the \textsc{bbody} component tends to zero and therefore that such a component would not be necessary anymore. 
The \textsc{phabs} N$_{H}$ value of 7.4$^{+1.0}_{-0.9}$$\times$10$^{20}$\,cm$^{-2}$ is consistent with the ISM value while the complex absorption via \textsc{pwab} indicates a N$_{H,max}$ of at least 10$^{24}$\,cm$^{-2}$ with $\beta$ equal to -0.90$\pm$0.01. Another difference with respect to the values inferred from the other spectra is the substantially lower absorption depth associated with the Oxygen edge, of 0.21$\pm$0.05. 
 We then assumed the O abundance as a free parameter during the fit, and also the Fe  which was tied to the Ni abundance. The fit resulted in Z$_{N}$\,=\,16.9$^{+2.3}_{-2.1}$\,Z$_{\odot}$, Z$_{O}$\,=\,1.3$^{+0.2}_{-0.1}$\,Z$_{\odot}$, and Z$_{Fe}$\,(=\,Z$_{Ni}$)\,=\,0.19$\pm$0.02\,Z$_{\odot}$, with $\chi^2_{red}$\,=\,1.19. 
In this scenario, the absorption from the \textsc{phabs} is less than 10$^{20}$\,cm$^{-2}$, with the \textsc{pwab} component indicating a N$_{H,max}$ of at least 10$^{24}$\,cm$^{-2}$ and $\beta$ equal to -0.79$\pm$0.01. 
The temperature from the \textsc{vbapec} component is $k$T\,=\,0.50$^{+0.1}_{-0.2}$\,keV ($k$T$_{max}$ in \textsc{vmcflow} was kept fixed to 43.4\,keV as it was derived from the XMM+NuSTAR for the same model).
The improvement with this model comes from a better description of the continuum around 2-5\,keV but it degrades the description of the ionized iron lines around 6.7\,keV and also fails to explain  the X-rays in the 0.55-0.7\,keV and 1-1.5\,keV energy ranges. The addition of another thermal plasma component did not improve the fit. 
Again, as in the analysis presented in Section \ref{sect:V2731_spctevol}, the \textsc{phabs} results in a unrealistically low column density. The results are summarized in Table \ref{tbl:v2731highnitrogen}.

The model has the OVII edge at 0.74\,keV, and some of the oxygen ions that absorbed such photons would emit the OVII He-like triplet. We then added a Gaussian line to the variable abundance model and its energy centroid converged to 0.58\,keV, the expected value of the OVII triplet. It results in $\chi^{2}_{\nu}$\,=1.15 but this time not only with low column density from the \textsc{phabs} component but understimating the spectra below 0.6\,keV. We them added the \textsc{bbody} component, resulting in a better description. The  continuum below 0.6\,keV is well described, the column density from \textsc{phabs} converges to $\sim$3.8$\times$10$^{21}$\,cm$^{-2}$, and the temperature of the \textsc{bbody}, contrary to the model presented in Section \ref{sect:V2731_spctevol}, has a low temperature of $\sim$\,28\,eV but an unrealistic normalization of 2.5. The $\chi^{2}_{\nu}$\, is 1.16. 

Because of the excess observed around 0.90-0.94\,keV, which is consistent with the position of the NeIX triplet, we add a new Gaussian component. The centroid energies of the Gaussian lines converge to 0.58$\pm$0.01\,keV and 0.94$\pm$0.01\,keV in good agreement with what is expected for the OVII and NeIX He-like triplets, respectively. 
For the blackbody component, we derive $k$T\,=\,39$^{+5}_{-4}$\,eV and normalization of (4.3$^{+4.0}_{-2.4}$)$\times$10$^{-2}$. 
The column from the \textsc{phabs} component is 2.9$^{+0.2}_{-0.3}$$\times$10$^{21}$\,cm$^{-2}$, consistent with the value expected from the ISM in the light of sight of V2731\,Oph. The fit also results in N$_{H,max}$\,=\, 68$^{+15}_{-20}$$\times$10$^{22}$\,cm$^{-2}$       and $\beta$\,=\,-0.77$^{+0.02}_{-0.01}$ for \textsc{pwab}, and $k$T\,=\,0.49$\pm$0.02\,keV for \textsc{vapec}. For the \textsc{vmcflow} component, we kept  $k$T$_{max}$\,=\,47.6\,keV. 
Also the fit, with $\chi^2_{red}$\,=\,1.12, resulted in 
Z$_{N}$\,=\,69$^{+16}_{-10}$\,Z$_{\odot}$, Z$_{O}$\,=\,4.6$^{+0.8}_{-0.5}$\,Z$_{\odot}$, 
and Z$_{Fe}$\,(=\,Z$_{Ni}$)\,=\,0.6$^{+0.1}_{-0.1}$\,Z$_{\odot}$. 
The best fit spectral parameters are presented in Table \ref{tbl:v2731highnitrogen}. In Fig. \ref{fig:v2731Suzaku} we present the fits from M6 and M9 for Suzaku/XIS1 (lines red and black, respectively) where are also presented the XIS0 and XIS3 spectra, and the corresponding results from M9.

We applied the above models to the XMM-{\it Newton} EPIC spectra. The fit with variable Z$_N$ was found to be statistically worse than the previous description using model M6 (Table \ref{tbl:v2731spctparam}), with $\chi^{2}_{\nu}$ of 1.14. When both Z$_{N}$ and Z$_{O}$ are allowed to be free, without the O absorption edge, results in difficulties to explain the emission from 0.7 to 0.8\,keV. Finally, we applied M8 (see Table \ref{tbl:models}) to the combined XMM-{\it Newton} and 20--75\,keV NuSTAR data, and obtained a  $\chi^{2}_{\nu}$ of 1.08 (compared to 1.01 for M6) with similar parameter values when exploring only the XMM-{\it Newton} spectra, particularly $k$T$_{max}$, which shows no statistically significant changes. The results are shown in Table \ref{tbl:v2731highnitrogen}. The addition of a \textsc{bbody} and the two \textsc{gaussian} line as in M9 do not improve the fit. With a low temperature ($k$T\,$\sim$\,15\,eV), the \textsc{bbody} component converges to a normalization of only $\sim$\,10$^{-19}$ meaning that such an addition is not supported from the data. A \textsc{gaussian} component at 0.94\,keV, as applied to the Suzaku data, is not constrained by the fit and fixing its parameters to the values derived from the Suzaku spectra does not improve the fit. Also, although still in agreement with the XMM-{\it Newton} data, the \textsc{gaussian} line around 0.58\,keV results in no significant improvement in the fit, still with $\chi^{2}_{\nu}$ of 1.08. The values derived from M8 (Table \ref{tbl:v2731highnitrogen}) are mutually consistent with those when applying M9 to the 0.3--10\,keV XMM-{\it Newton} plus the 20-75\,keV NuSTAR spectra althought these fits are not shown.

Thus, even though we have not 
found a completely satisfactory fit to the soft X-ray spectrum, our idea to combine XMM-{\it Newton} and 20--75\,keV NuSTAR data appears to be valid. First, the spectral energy distribution for hard X-rays in both set of data are mutually consistent -- and confirmed from the whole spectral range covered by NuSTAR (Section \ref{sct:V2731Oph_Nustar}). 
The difficulties with either model in describing the Suzaku data with all applied models, and the XMM-{\it Newton} spectra with the alternative M8 and M9 models, are limited to the soft end of the spectrum ($<$1\,keV). 
We notice that the elemental abundances derived independently from the Suzaku and XMM-{\it Newton} spectra do not converge, both indicate that anomalies are present at least for the N, O, and Fe. 
Second, the absorbed and unabsorbed 20-75\,keV flux differ by less than 5\%, showing that the absorption such as detected in the system, even though it is very high, has no significant rule in this spectral region. 
In addition, the investigation revealed that the luminosity of the source at 20--75\,keV during the XMM-{\it Newton} observation is $\sim$\,35\% less than that observed during the NuSTAR observation, with the unabsorbed flux varying from 
$\sim$\,2.5$\times$10$^{-11}$\,erg\,cm$^{-2}$\,s$^{-1}$ to $\sim$\,3.9$\times$10$^{-11}$\,erg\,cm$^{-2}$\,s$^{-1}$ (over 9 years). 
The estimated unabsorbed 20--75\,keV flux from both Swift and Suzaku datasets is $\sim$\,2.1$\times$10$^{-11}$\,erg\,cm$^{-2}$\,s$^{-1}$ (over 2 years).
Since the soft X-ray complexity is limited to energies below 0.8\,keV, we will proceed by returning to the spectral model of \citet{2008A&A...481..149D} in the analysis of hard X-ray data, and comment on the soft X-ray complexity in Section \ref{sct:v2731complex}.

\subsubsection{Spectroscopy using the full NuSTAR data}
\label{sct:V2731Oph_Nustar}

The variability of V2731\,Oph in X-rays prevents the use of non-contemporaneous campaigns to carry out a simultaneous spectral fit. 
Therefore, we instead explored the full spectral range covered by NuSTAR (3--75\,keV) initially using models M6. As expected, the \textsc{phabs} (that goes virtually to zero; N$_{H}$\,=\,10$^{13}$\,cm$^{-2}$), \textsc{bbody}, \textsc{apec}, and \textsc{edge} components do not contribute to the fit and they were removed from the model. 
A fit that keeps the N$_{H,max}$ of the \textsc{pwab} component fixed to 156$\times$10$^{22}$\,cm$^{-2}$ and the $k$T$_{hot}$ of the \textsc{mkcflow} to 47.6\,keV, as derived from the analysis of the XMM-{\it Newton} and 20--50\,keV NuSTAR spectra, yields $\chi^{2}_{\nu}$\,=\,1.48. 
Freezing the thermal component to 47.6\,keV and allowing the N$_{H,max}$ to vary during the fit results in a good description of the spectra with N$_{H,max}$\,=\,753$^{+60}_{-62}$$\times$10$^{22}$\,cm$^{-2}$ and $\beta$\,=\,-0.68$^{+0.01}_{-0.01}$, for $\chi^{2}_{\nu}$\,=\,1.00 (Table \ref{tbl:v2731spctparam}). The abundance was kept fixed to 0.30\,$Z_{\odot}$. Similar values were obtained when allowing the high temperature of the \textsc{mkcflow} varying during the fit.
 kT$_{max}$ of 47.6$^{+4.2}_{-4.5}$\,keV derived from the joint NuSTAR-XMM-{\it Newton} fit suggests a WD mass of 0.92$\pm$0.04 M$_\odot$, if the shock temperature reflects the gravitational potential just above the white dwarf surface \citep[following][]{1973PThPh..49.1184A}.

\subsubsection{Investigating the hypothesis of reflection}

We move back to the spectral investigation of V2731\,Oph to verify how the occurrence of X-ray reflection in the system would affect the results. Contrary to the case of TV\,Col, it was not possible to constrain the reflection - if it is present - for V2731\,Oph. Adding the \textsc{reflect} component and allowing the reflection scaling factor to vary during the fits resulted, systematically, in values greater than 3, which are unrealistic. We froze this value to 1 while keeping cos$i$\,=\,0.45 and see how this conservative assumption would impact the spectral results when reflection is present. In doing this, we followed the same procedure as in the case without \textsc{reflect}: we investigate the EPIC XMM-{\it Newton} and the 20--75\,keV NuSTAR (FPMA\,+\,FPMB) simultaneously, then individually the Swift/XRT, the Suzaku (XIS0+XIS3 and XIS0+XIS1+XIS3), and the 3--75\,keV NuSTAR spectra.
The results are summarized in Table \ref{tbl:v2731spctparam}.
$k$T$_{max}$ of 55.1$^{+6.2}_{-4.4}$\,keV derived from the joint NuSTAR-XMM-{\it Newton} fit (with the abundance fixed to 0.3\,Z$_\odot$) suggests a WD mass of 0.99$^{+0.05}_{-0.03}$ M$_\odot$, if the shock temperature reflects the gravitational potential just above the white dwarf surface.

The inclusion of reflection does not change at the 1\,$\sigma$ level the high temperature in the \textsc{mkcflow} model. But this result can be masked by effects of the photoelectric absorption, that is not well constrained from the NuSTAR data alone. A conclusive investigation of reflection in V2731\,Oph demands high signal-to-noise spectroscopy covering at least the 0.3--75\,keV energy range. 
High-resolution spectroscopy would play a crucial role in the description of the lines in the soft X-rays and therefore of the continuum and, finally, of the intervening absorber. As for now, our results suggest that 35\% of the total flux at 10-50\,keV derived for V2731\,Oph from the XMM-{\it Newton} fit can be due to reflection -- similar to the fraction inferred for TV\,Col, in which the presence of reflection is clear.

We estimate from the XMM-{\it Newton} (pn; on 2005/08/29) observation an EW of 145$\pm$40\,eV for the fluorescent iron line at 6.4\,keV. The NuSTAR data have a poorer spectral resolution but suggest a lower value on 2014/05/14 of about 69\,eV even in a situation in which X-rays seems to be more absorbed than they were during the XMM-{\it Newton} observation (Figure \ref{fig:v2731ophspct}, inset). Perhaps the low equivalent width suggests that the 6.4\,keV line can be accounted only by the absorbers with only a small contribution, if any, from reflection.

\subsubsection{X-ray luminosity and mass accretion rate}

V2731\,Oph is a luminous IP in the X-rays, with a high mass accretion rate. Here we concentrate on results from the XMM-{\it Newton} and NuSTAR analyses. From the cooling flow model, we infer directly from its normalization parameter value that the mass accretion rate during the XMM-{\it Newton} observation (2005-08-29) was 5.1$^{+2.7}_{-0.7}$$\times$10$^{-9}$ M$_{\odot}$\,yr$^{-1}$, with a bolometric luminosity of 4.2$\times$10$^{35}$($d$/2300 pc)$^{2}$\,erg\,s$^{-1}$, while the rate during the NuSTAR observation (2014-05-14) was (8.4$\pm$0.2)$\times$10$^{-9}$ M$_{\odot}$\,yr$^{-1}$. Because NuSTAR does not cover energies below 3\,keV, where the system has a significant contribution from additional components which are therefore not constrained from those data, we cannot infer the X-ray bolometric luminosity securely from the NuSTAR data. To carry out a more realistic comparison, we adopt the luminosity in the 3--75\,keV band. The values are 4.4$\times$10$^{34}$($d$/2300 pc)$^{2}$\,erg\,s$^{-1}$ and 7.8$\times$10$^{34}$($d$/2300 pc)$^{2}$\,erg\,s$^{-1}$ for the XMM-{\it Newton} and NuSTAR observations, respectively. 
We estimate that the 3-75\,keV luminosity is about 10\% less than the bolometric X-ray luminosity.

      \section{Timing properties from the NuSTAR observations}

The NuSTAR observations allowed us to investigate the spin variability of both TV\,Col and V2731\,Oph in hard X-rays. 
We proceeded from background subtracted light curves in three energy bands: 3--6 keV, 6--10 keV, and 10--30 keV. 
We chose 60 s bins for TV Col, given its long spin period (resulting
    in 32 bins per cycle), and 10 s bins for V2731 Oph, given its short
    period (13 bins per cycle).
Since the NuSTAR count rates of these IPs are modest, we neglected deadtime correction, which should be small. We also neglected barycentric correction to simplify the analysis procedure, since we intended to limit our analysis to the times of the NuSTAR observations. We combined the FPMA and FPMB light curves in the three bands and searched for periodic modulations using Fourier power spectrum and the normalization according to \citet{1982ApJ...263..835S}. 
 There is no Swift data at the time of the NuSTAR observation of
    V2731 Oph, while that contemporaneous with the NuSTAR observation
    of TV Col is too short to provide conclusive results.
 As timing analysis of all
    archival X-ray data is beyond the scope of our work, given that extensive body of
    work already exists in print, we focus on the NuSTAR observations.

\begin{figure}
\centerline{
\includegraphics[angle=-0,scale=0.53]{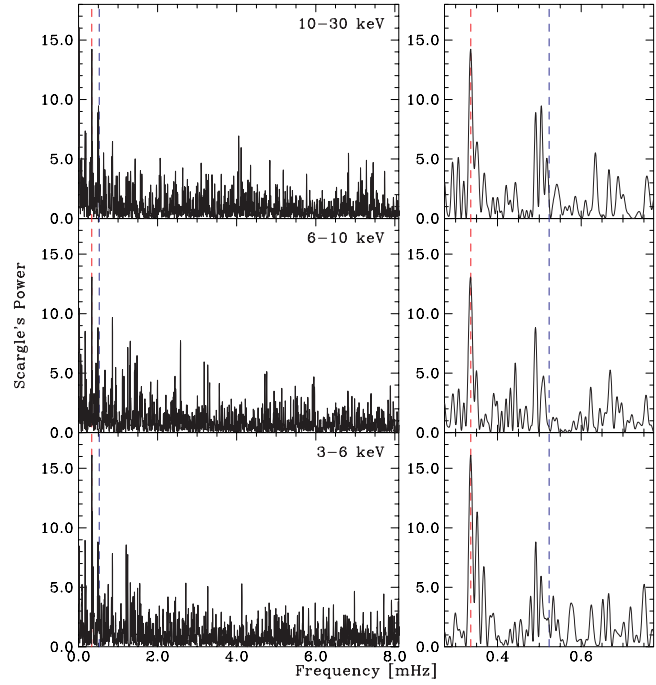}
}
\caption{TV\,Col: periodogram from NuSTAR data in three X-ray energy bands; 3-6\,keV (bottom), 6-10\,keV (middle) and 10-30 keV (top).
The set of panels show the power spectra over the range of periods
we investigated (longer than 120\,s). The blue dashed line is drawn
at the known spin period of TV\,Col, 1909.7\,s, while the red dashed
line indicates the prominent peak at 2976\,s. The right set of panels
show the power spectra around 1909.7\,s in more detail.
\label{fig:tvcol_scargle}}
\end{figure}

\begin{figure}
\centerline{
\includegraphics[angle=-0,scale=0.53]{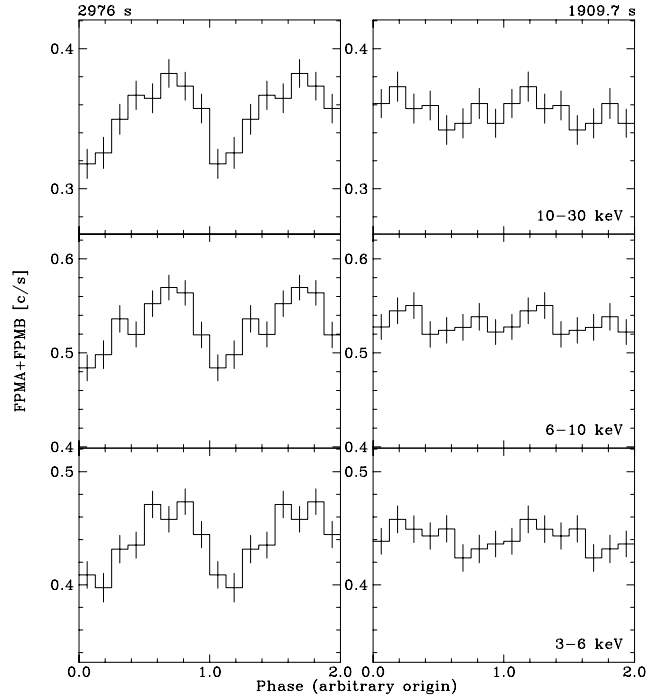}}
\caption{TV\,Col: folded light curves (on the prominent period of 2976 s, left, and on the known spin period of 1909.7 s, right) from NuSTAR data in three X-ray energy bands; 3-6\,keV (bottom), 6-10\,keV (middle) and 10-30 keV (top). \label{fig:tvcol_lc}}
\end{figure}

\subsection{TV\,Col}

The known spin modulation in TV\,Col, refined by \citet{2004AJ....127..489R} as 1,909.7$\pm$2.5\,s, was not detected in the NuSTAR data. Figure \ref{fig:tvcol_scargle} shows the power spectra up to 8 mHz from the three energy ranges cited above, where the blue vertical dashed lines mark the position of the expected spin frequency. The red lines in the figure mark the highest peak below 10\,keV, at $\sim$\,2,976\,s, while it is the second in power in the 10--30\,keV energy range after the one at very low frequency ($\sim$\,96,000\,s). The peak at 2,976\,s is statistically significant if white (i.e., frequency independent) noise is assumed. However, no previous reports of a $\sim$3,000 s period exist, so we consider this to be a noise peak.

\begin{figure}
\centerline{
\includegraphics[angle=-0,scale=0.53]{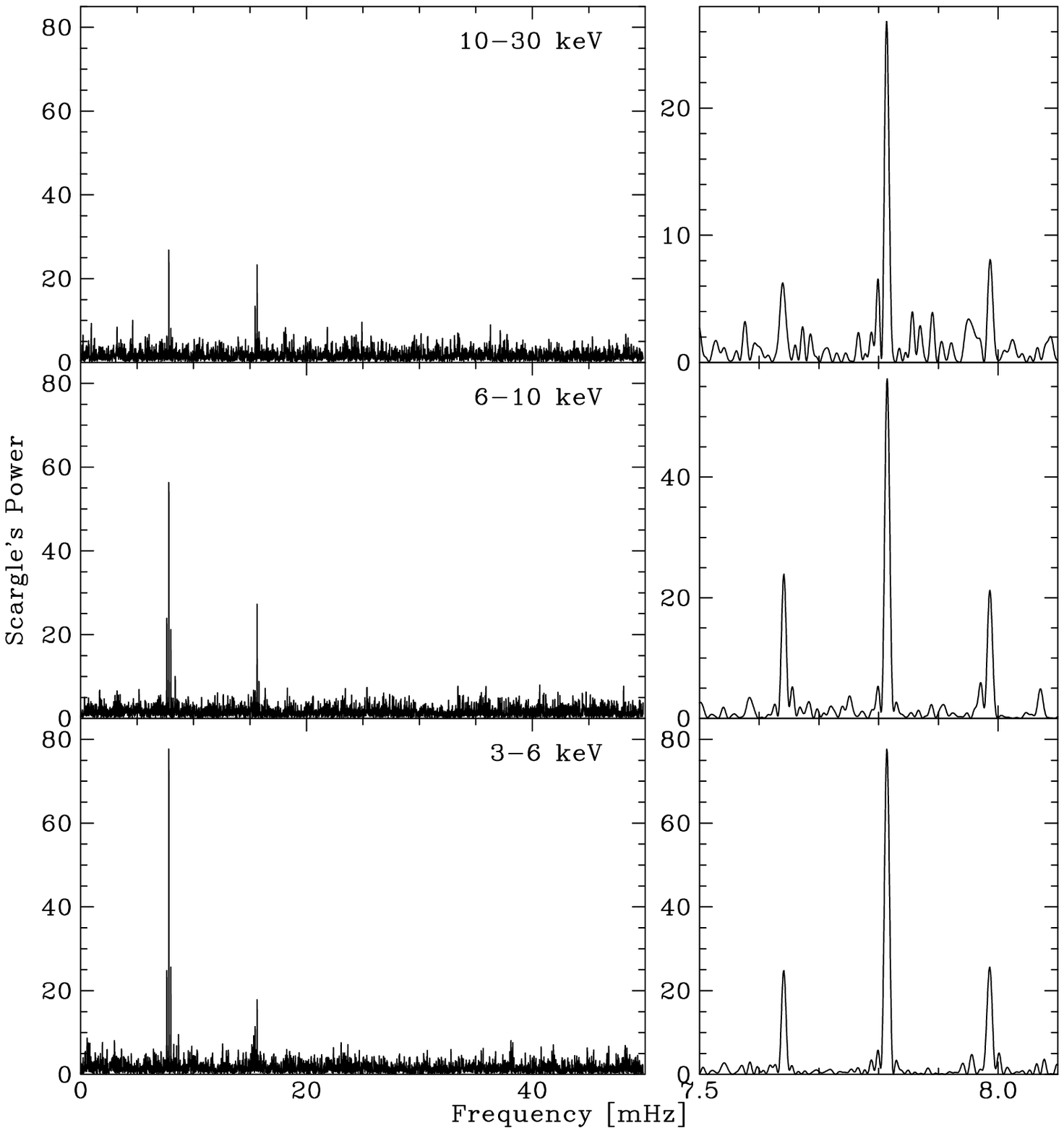}}
\caption{V2731\,Oph: periodogram from NuSTAR data in three X-ray energy bands; 3-6\,keV (bottom), 6-10\,keV (middle) and 10-30 keV (top).
The set of panels show the power spectra over the range of periods
we investigated (longer than 20\,s). The right set of panels show
the power spectra around 128\,s in more detail (note the different
vertical scales).
\label{fig:v2731_scargle}}
\end{figure}

\begin{figure}
\centerline{
\includegraphics[angle=-0,scale=0.53]{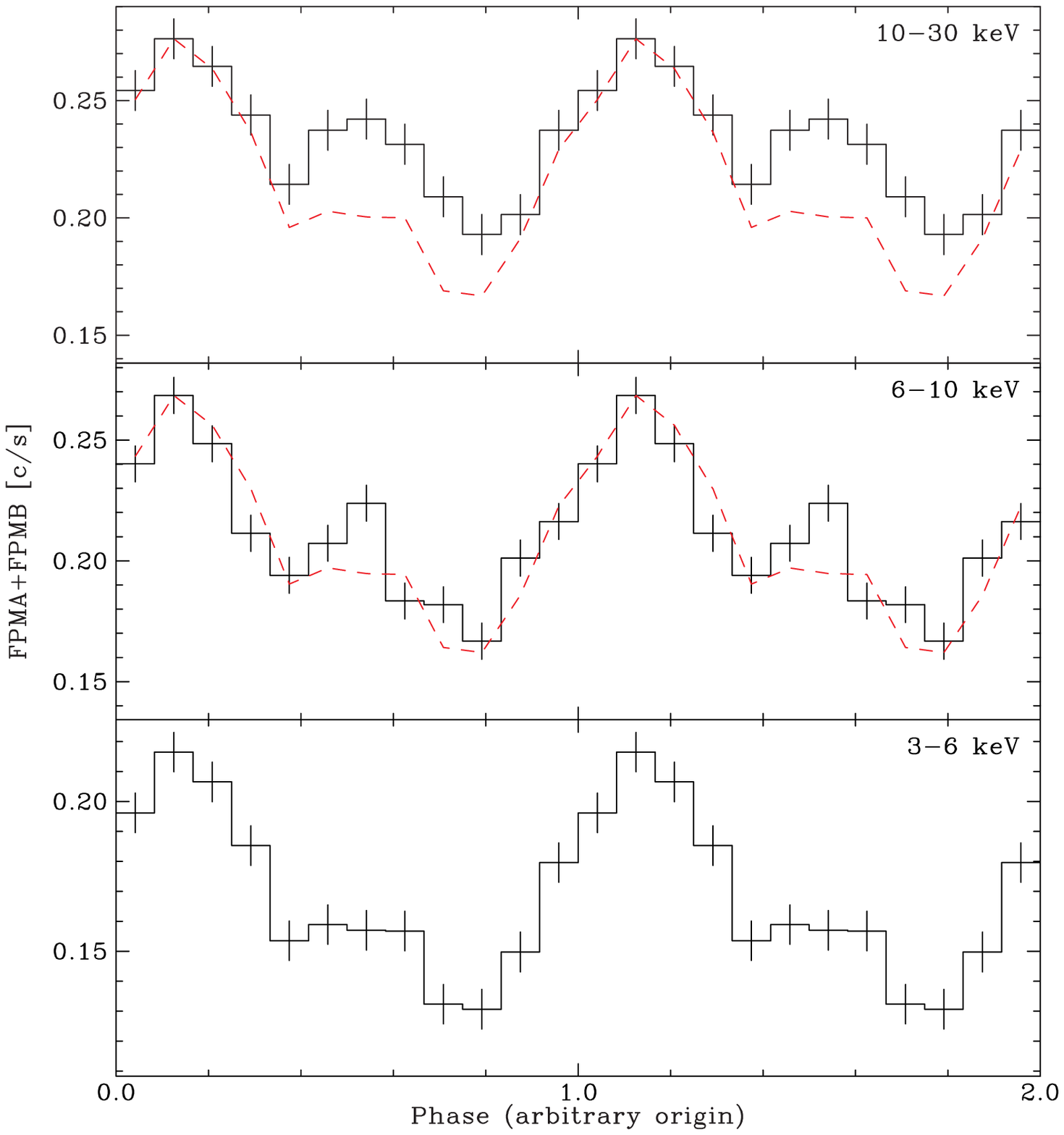}
}
\caption{V2731\,Oph: light curves folded on the 127.98\,s spin
period from NuSTAR data in three X-ray energy bands; 3-6\,keV (bottom), 6-10\,keV (middle) and 10-30\,keV (top). \label{fig:v2731_lc}}
\end{figure}

We show folded light curves with the periods of 1,909\,s and 2,976\,s in the left and right panels of Fig. \ref{fig:tvcol_lc}, respectively.
As anticipated from the power spectrum, no modulation is associated with the spin period.
Using amplitude/average convention, the relative modulation amplitudes for the 2,976 s peak are
 $\sim$\,7.5\% (3-6 keV), $\sim$\,6.6\% (6--10 keV), and 8.4\% (10--30 keV), and they can be taken as
conservative upper limits of the spin (1,909.7\,s) signal.
In comparison, \citet{2004AJ....127..489R} found spin amplitudes of 14\% (2-5 keV), 8.5\% (5-10 keV), and 6\% (10-20 keV). Such modulations would have been detectable in the NuSTAR data below 10 keV, if not above. Thus, we conclude that the X-ray spin amplitude of TV Col is variable and that we caught it in a low spin modulation state.
We note that optical observations of TV\,Col have long indicated
    variable spin modulation amplitude \citep[see, e.g.,][]{1985A&A...143..313B,1988MNRAS.233..759B}.

\subsection{V2731\,Oph}

The 128\,s spin period of V2731 Oph \citep{2008A&A...481..149D} and the first harmonic are
clearly dominant in the periodogram presented in Fig. \ref{fig:v2731_scargle}.
The left panels show the periodogram over the entire frequency range (0-50 mHz,
from the light curves with time bin of 10\,s) in the three X-ray energy bands under investigation. All the three left panels have the same Y scale. 
The right panels show the expanded view around the fundamental frequency (with variable Y scales), where the peak corresponding to the 128\,s period is seen and its two 1-cycle-per-spacecaft-orbit aliases. 
The spin period for V2731\,Oph is estimated to be 127.981$\pm$0.022\,s during the NuSTAR observation, consistent with the previous determination in the optical
    \citep[e.g., 127.999909(49)\,s;][]{2005MNRAS.361..141G}
    and X-rays \citep[e.g., 128.02$\pm$0.02\,s;][]{2008A&A...481..149D}.

The folded light curves are not quite sinusoidal, although
mostly so in the 3--6 keV range (Fig. \ref{fig:v2731_lc}). The red dashed lines in the upper and middle panels are copies of the 3--6 keV spin profile, showing that the main peak remained the same, but a secondary
peak is very pronounced in the 10--30 keV range.

\section{Discussion}

\subsection{Soft X-ray complexity of V2731\,Oph}
\label{sct:v2731complex}

The fact that V2731\,Oph has a complex X-ray emission was already reported by \citet{2008A&A...481..149D}. Our analysis show that their model cannot fit the Suzaku XIS1 spectrum. Moreover, a blackbody temperature of 0.21 keV and 0.18 keV, as derived from Suzaku XIS0+XIS3 and XIS0+XIS1+XIS3 spectra, respectively, is clearly unphysical for WD accretion. Considering the local Eddington limit, the highest effective temperature for a black body contribution would be $\sim$140\,eV for a 1.4\,M$_{\odot}$ WD \citep{1987MNRAS.226..725W}, which would imply a shock temperature $kT_{max}$ of 230\,keV. Even the 94\,eV temperature, as derived from XMM-{\it Newton} data and consistent with Swift observations, demand a WD mass of at least 1.2\,M$_{\odot}$, with a high specific accretion rate of about 200\,g\,cm$^{2}$\,s$^{-1}$ and $k$T$_{max}$\,$>$\,100\,keV, and therefore inconsistent with our spectral analysis.

Based on the Suzaku XIS1 spectrum, we have proposed an alternative spectral model that does not include a hot blackbody. For other instruments, but still acceptable for the Suzaku XIS0 and XIS3 spectra,
this model results in a poorer fit than was obtained with the model with a blackbody. One plausible explanation is the limitation of the {\tt pwab} model. While this model is the most sophisticated in terms of reproducing the likely distribution of the number of lines of sight as a function of N$_H$ \citep{1998MNRAS.298..737D}, it assumes that the absorber has the solar abundance and is not ionized. If the emitting plasma is overabundant in nitrogen, so, too, should the absorbing plasma, which is not currently taken into account. Moreover, the presence of the OVII edge means the latter assumption is invalid. We currently model this using an edge model in addition to {\tt pwab}, but we believe this is not the correct model for this situation. In reality, the absorption below 0.78 keV should be less than what {\tt pwab} predicts, because the complex absorber is ionized and is mostly transparent at these energies. We therefore believe that the current inadequecies of the model is the likely explanation for the imperfect fit. It is likely that high-resolution, high signal-to-noise X-ray spectroscopy combined with an improved grid of models would be needed to see if the alternative spectral description is correct. In the mean time, the abundance of N should be investigated using optical and UV spectroscopy.
 Nitrogen overabundance has been seen in a number of CVs, both magnetic
    and non-magnetic \citep[see, e.g.,][]{2003ApJ...594..443G,2012ApJS..199....7F}, and indicates the presence of
    CNO-processed matter in the system. This, in turn, can be either due
    to prior nuclear evolution of the donor \citep{2002MNRAS.337.1105S} or pollution by previous nova eruptions
     \citep{1998MNRAS.301..699M}.

For the main purposes of this work, the exercise we carried out by using three sets of X-ray observations for V2731\,Oph shows that: (i) while there is no significant evidence of change in the spectral distribution of the hottest thermal component, its luminosity is variable and its X-ray photons suffer the effect of a variable absorbing column, (ii) the hottest thermal component is in agreement with a cooling flow, stratified multi-thermal model which is expected for IP systems, (iii) soft X-ray spectrum is variable in flux and in spectral distribution. 
 Given these findings, we believe our results are robust
    regarding the complex absorber, reflection amplitude, and the
    maximum temperature of the post-shock region, despite our
    imperfect understanding of the soft ($<$1\,keV) spectrum of
    V2731 Oph.

\subsection{Clues from the plasma temperature and reflection amplitude}
\label{sct:tandrefl}

While the maximum temperature ($k$T$_{max}$) derived from the \textsc{mkcflow} model traces the gravitational potential well at the accretion column shock, the reflection amplitude ($\Omega_{r}$) depends on the shock height ($h$) with respect to the stellar surface. Thus, these two spectral signatures allow us to put constraints on the white dwarf mass and on the accretion geometry of the system.

For a point-like emission region placed at height $h$ above the white dwarf, with $h$ expressed as a fraction of the stellar radius, the stellar surface covers the solid angle \citep[see Figure 2 of][]{2018PASJ...70..109T}:

\begin{equation}
\Omega\,=\,2\pi (1 - \sqrt{1-1/(1+h)^2})
\label{eq:omega}
\end{equation}

The reflection amplitude is defined as the solid angle covered by the reflector as seen from the emitter in such a way that it is 1.0 when the solid angle subtended is 2$\pi$ (half the sky). This condition corresponds to $h$\,=\,0 in Eq. \ref{eq:omega}, and therefore $\Omega_{r}$\,$\equiv$\,$\Omega$/2$\pi$. Thus, a reflection amplitude of 0.88$\pm$0.13 as derived from the spectral fit of TV\,Col (see Table \ref{tbl:spct_tvcol}) requires $h$ of order 0.7\% of the white dwarf radius. Such a small shock height implies that $k$T$_{max}$ traces the gravitational potential well virtually at the WD surface and therefore the white dwarf mass determined from our spectral fit (0.735$\pm$0.015 M$_\odot$) does not require additional corrections.

On the other hand, as first derived by \citet{1973PThPh..49.1184A}, the geometrical parameter $h$ is physically determined by the
      post-shock cooling time ($t_{c}$), that defines the remaining time for the accreted material to settle onto the white dwarf. Higher density ($\rho$) in the accretion column accelerates the cooling ($t_{c}$\,$\propto$\,1/$\rho$), resulting in smaller $h$. Thus, $h$ is a function of white dwarf mass and specific (or per unit area) accretion rate (g\,s$^{-1}$\,cm$^{-2}$). 

      Following \citet{1973PThPh..49.1184A}, the low $h$ derived from the reflection in TV\,Col ($\sim$\,0.7\%) is due to a 
       high specific accretion rate, of order 15\,g\,s$^{-1}$\,cm$^{-2}$. 
       Comparing the specific accretion rate with the mass accretion rate derived from the \textsc{mkcflow} model, of (3.8$\pm$0.2)$\times$10$^{-10}$ M$_{\odot}$\,yr$^{-1}$, we estimate that the fractional area over which
      accretion happens ($f$) is of about 0.02\% for TV Col: a relatively high specific accretion rate, and a small accretion spot.
        The best estimate for the spot size in IPs is that for
    XY Ari, for which \citet{1997MNRAS.291...71H} used the X-ray
    eclipses to arrive at an estimate of $<$\,0.002. If reliable
    determination of reflection amplitude becomes routine, there
    is a potential to repeat the kind of analysis we have performed
    for TV Col to estimate the spot sizes of IPs. This might allow
    future researchers to perform a statistical study against other
    system parameters.
      
On the other hand, we did not detect an unambiguous sign of reflection in the NuSTAR observation of V2731\,Oph. Thus we are unable to derive the specific accretion rate or the shock height in V2731\,Oph from its X-ray spectrum.

\subsection{Additional constraints from the photometry}

X-ray spin modulation of IPs is usually due to complex absorption \citep{1989MNRAS.237..853N}, which is not expected to produce significant spin modulation above 10\,keV. Therefore, when spin modulation is seen above 10\,keV, a geometrical explanation is required. A likely scenario is that non-negligible shock height ($h$\,$>$\,0.1) results in both poles being visible over a range of viewing geometry \citep{2001A&A...377..499D}.

The lack of detectable spin modulation in the NuSTAR data on TV\,Col can be explained as follows. First, we do not expect geometrical spin modulation because the shock in TV\,Col is very close to the stellar surface ($h \sim$0.7\%). Secondly, the maximum energy at which we can detect X-ray spin modulation due to complex absorption) depends on N$_{H,max}$. Our spectral fit shows N$_{H,max}$ of only $\sim$\,10$^{23}$\,cm$^{-2}$ during the NuSTAR observation, so the effects of complex absorption are mostly confined to photon energy below $\sim$3\,keV. The complex absorption appears to be variable from epoch to epoch according to our analysis of the Swift data, which also extends to lower energies: however, these Swift data do not have sufficient spin phase coverage and our timing analysis was inconclusive. We expect that X-ray observations which showed spin modulations above 3 keV were obtained when TV\,Col was in a higher N$_{H,max}$ state.

In V2731\,Oph, N$_{H,max}$ is high, so the low energy ($<$\,10 keV) spin modulation may well have a contribution due to complex absorption. Nevertheless, the clear detection of the spin period above 10\,keV (Fig. \ref{fig:v2731_scargle}) implies a strong geometrical component to the X-ray spin modulation in this IP. This indicates a non-negligible shock height for V2731\,Oph. We thus expect a low reflection amplitude, which can explain the absence of evidence of reflection in the X-ray spectrum of the system. In that sense, V2731\,Oph appears similar to V709\,Cas \citep{2015ApJ...807L..30M}.

At first sight, it might seem surprising that a system as luminous V2731\,Oph should have a non-negligible shock height $h$. We suggest two factors which may explain the large $h$. First, recall that $h$ is determined, not by the total accretion rate, but by the specific accretion rate. We can explain the non-negligible $h$ by invoking a large $f$. A major factor controlling $f$ is the pinching by the dipole magnetic field (B\,$\propto$\,$r^3$), the degree of which is determined by the magnetospheric radius. If we assume approximate spin equilibrium, V2731\,Oph (P$_{spin}$\,=\,128\,s) must have a much smaller magnetosphere than TV\,Col (P$_{spin}$\,=\,1,911\,s), so dipole pinching is much less effective.  Secondly, V2731\,Oph may well have a massive WD. The free-fall velocity is higher, which reduces the post-shock density, which reduces the X-ray cooling efficiency. Since the post-shock temperature is also high, this increases $t_c$, and hence $h$.

\subsection{Our targets and the standard evolutionary scenario}

The mass transfer in CVs with orbital period less than $\sim$10 hrs
are driven by angular momentum loss, and is expected to be of order
3-5$\times$10$^{-9}$ M$_{\odot}$\,year$^{-1}$ for a 5.5 hr period CV, 
as appropriate for TV\,Col
\citep{2011ApJS..194...28K}. Both the normalization of the \textsc{mkcflow} model
and the bolometric X-ray luminosity of TV\,Col suggests a lower
accretion rate, unless a large amount of transferred material is
not accreted and is lost, perhaps propellered out of the system, therefore reducing the expected accretion rate and consequent X-ray emission. The observed X-ray luminosity is typical of IPs \citep{2018A&A...619A..62S}.

This standard evolutionary scenario does not apply to V2731\,Oph,
and so we have no quantitative prediction for the mass transfer
rate in this system. The BAT band luminosity of V2731\,Oph is already
log\,L$_{X}$(erg\,s$^{-1}$)\,=\,34.6 \citep{2018A&A...619A..62S}, the highest of all IPs, 
 and the uncertain evolutionary driver must be capable of sustaining
    a high mass transfer rate. 
The overabundance of N, if confirmed, would place V2731\,Oph among
the subset of CVs known to be overabundant in N 
\citep[see, e.g.,][]{2003ApJ...594..443G}. This would imply that the donor matter has been
CNO-processed: the secondary of V2731\,Oph may be evolved in the
normal (single star evolution) sense, or it may indicate that this is
a post thermal-timescale mass-transfer (TTMT) system \citep{2002MNRAS.337.1105S}.

\subsection{White dwarf masses}
\label{sct:wdmass}

We have shown that both a complex absorber and reflection are necessary to fit the wide-band X-ray spectrum of TV\,Col. The combined NuSTAR+Swift data allowed us to constrain both, as well as the maximum temperature of the X-ray emitting plasma. 
The high reflection amplitude supporting a small shock height $h$ and the long spin period indicate that the inner radius of the truncated disk is large. Taking the maximum plasma temperature as a measure of the gravitational potential just above the WD surface, we determine the WD mass to be 0.735$\pm$0.015 M$_\odot$. Our value is somewhat smaller than, but consistent with, the previous NuSTAR measurement by \citet{2016ApJ...826..160H}.

On the other hand, we were unable to constrain the reflection amplitude in V2731\,Oph. This has several effects on the mass measurement. First, our derived values for $kT_{max}$ depends on whether we include reflection (with amplitude fixed to 1.0) or not. Also, our values, particularly the case without reflection (0.92$\pm$0.04 M$_\odot$) is significantly lower than what \citet{2016ApJ...826..160H} derived; the reason for this is unclear. Moreover, we were unable to constrain the shock height, but the high energy ($>$10 keV) spin modulation suggests it is not small. Because of this, and because V2731\,Oph is a fast rotator that probably requires correction for small inner radius of the disk, we consider our numbers to be lower limits for the true WD mass in this system.

\section{Summary}

We have combined analyses of broad-band X-ray spectroscopy and photometry, as in \citep{2015ApJ...807L..30M}, to study two IPs, TV\,Col and V2731\,Oph, already analyzed by \citet{2016ApJ...826..160H}. We have gained valuable insights into the geometry and physics of accretion in these systems.

\begin{enumerate}
\item TV\,Col: 

\begin{itemize}
\item Complex absorption, such as described by the {\tt pwab} model is required to obtain a good spectral fit.
\item Reflection is also clearly detected, with an amplitude of 0.88$\pm$0.13, indicating a small shock height caused by high specific accretion rate.
\item We do not detect the spin modulation in the NuSTAR data, 
showing that the amplitude of the spin modulation can vary over different epochs, as is also seen in at optical wavelengths.
\item The white dwarf mass is determined to be 0.735$\pm$0.015 M$_\odot$.
    \end{itemize}

\item V2731\,Oph:
 
  \begin{itemize}
\item The X-ray spectrum is affected by a complex and variable local absorption, with a much higher N$_H$ than for TV\,Col.
\item The spectral model, which includes a hot blackbody, used to fit the XMM-{\it Newton} data does not work for Suzaku XIS1 data. We have tentatively proposed an alternative model without such a blackbody, but with an overabundance of nitrogen.
\item The NuSTAR data did not allow us to constrain the reflection amplitude. The detection of high energy ($>$10\,keV) spin modulation suggests a tall shock.
\item A combination of clues points to a high mass white dwarf in V2731\,Oph.
    
  \end{itemize}
\end{enumerate}

\acknowledgments

We thank the anonymous referee for numerous helpful suggestions.
R.L.O. was partially supported by the Brazilian agency CNPq (Universal Grants 459553/2014-3, PQ 302037/2015-2, and PDE 200289/2017-9). 
This work has made use of data from the European Space Agency (ESA) mission
{\it Gaia} (\url{https://www.cosmos.esa.int/gaia}), processed by the {\it Gaia}
Data Processing and Analysis Consortium (DPAC,
\url{https://www.cosmos.esa.int/web/gaia/dpac/consortium}). Funding for the DPAC
has been provided by national institutions, in particular the institutions
participating in the {\it Gaia} Multilateral Agreement.
This research has made use of the NuSTAR Data Analysis Software (NuSTARDAS) jointly developed by the ASI Science Data Center (ASDC, Italy) and the California Institute of Technology (Caltech, USA). This research has made use of the XRT Data Analysis Software (XRTDAS) developed under the responsibility of the ASI Science Data Center (ASDC), Italy. 

%

\vspace{5mm}
\facilities{NuSTAR, Swift, Suzaku, XMM-{\it Newton}.}



\end{document}